\documentclass[amsmath,prb,showpacs,twocolumn,superscriptaddress]{revtex4}

\usepackage{graphicx}

\begin{document}

\title{Inelastic tunneling effects on noise properties of
       molecular junctions}
\date{\today}
\author{Michael Galperin}
\affiliation{Department of Chemistry and Nanotechnology Center,
   Northwestern University, Evanston IL 60208}
\author{Abraham Nitzan}
\affiliation{School of Chemistry, The Sackler Faculty of Science,
   Tel Aviv University, Tel Aviv 69978, Israel}
\author{Mark A. Ratner}
\affiliation{Department of Chemistry and Nanotechnology Center,
   Northwestern University, Evanston IL 60208}

\begin{abstract}
The effect of electron-phonon coupling on the current noise in a molecular
junction is investigated within a simple model. The model comprises a
1-level bridge representing a molecular level that connects
between two free electron reservoirs and is coupled to 
a vibrational degree of freedom representing a molecular vibrational mode.
The latter in turn is coupled to a phonon bath that represents the
thermal environment. We focus on the zero frequency noise spectrum
and study the changes in its behavior under weak and strong electron-phonon
interactions. In the weak coupling regime we find that the noise
amplitude can increase or decrease as a result of opening of an inelastic 
channel, depending on distance from resonance and on junction asymmetry.
In particular the relative Fano factor decreases with increasing
off resonance distance and junction asymmetry.
For resonant inelastic tunneling with strong electron-phonon coupling 
the differential noise spectrum can show 
phonon sidebands in addition to a central feature. 
Such sidebands can be observed when displaying the noise against
the source-drain voltage, but not in noise vs. gate voltage plots
obtained at low source-drain bias.
A striking crossover of the central feature
from double to single peak is found for increasing asymmetry in
the molecule-leads coupling or increasing electron-phonon interaction.
These variations provide a potential diagnostic tool. 
A possible use of noise data from scanning tunneling microscopy experiments 
for estimating the magnitude of the electron-phonon interaction 
on the bridge is proposed.
\end{abstract}


\maketitle

\section{\label{intro}Introduction}
Development of molecular electronics as a complement to traditional
semiconductor-based electronics provides challenges both experimentally and 
theoretically.\cite{MEB,AN,Reed,Cuniberti,Bowler}
Early studies of molecular junctions were restricted to
measurement of current-voltage (or conductance-voltage) characteristics
of such junctions and their dependence on junction parameters such 
as wire length, molecular structure, molecule-lead coupling,
and temperature.\cite{lang,kubiak,reed_tour,heath,joachim,weber}
More recently, charging (Coulomb blockade) and inelastic and mechanical 
effects appear at the forefront of 
research.\cite{park,bjornholm,mceuen,natelson,ho,zhitenev}
In particular, inelastic electron tunneling spectroscopy (IETS)
has become an important diagnostic and characterization tool,
indicating the presence and structure of a molecule in the 
junction\cite{ho,zhitenev} and providing information on the junction 
structure.\cite{ruitenbeek}

In addition to the $I/V$ behavior, noise characteristics can provide
information about molecular junctions.\cite{noiserev_Buttiker}
Noise measurements in mesoscopic tunnel junctions have been
under study for a long fine, e.g. in
semiconductor double barrier resonant tunneling structures 
(DBRTS),\cite{e_DBRTS_Li,e_DBRTS_Safonov,e_DBRTS_Nauen,e_DBRTS_Jung}
quantum point contacts (QPC),\cite{e_QPC_Li,e_QPC_Washburn,
e_QPC_Liefrink,e_QPC_Reznikov} Coulomb blockaded Josephson 
junctions,\cite{e_JJ_Delahaye,e_JJ_Lindell}
and advances in understanding their diagnostic capabilities are being made
(see, e.g.~\onlinecite{e_3rdmoment_Reulet}). Results of measurements in
molecular tunnel junctions are expected to appear in the near future, 
though these measurements are difficult due to the background
$1/f$ noise caused by charge fluctuations in the environment.
Many theoretical studies of noise in nanojunction transport
are available in the literature
(see Ref.~\onlinecite{noiserev_Buttiker} and references therein). In particular
noise in DBRTS was studied within the nonequilibrium Green function (NEGF) 
formalism in the ballistic\cite{t_DBRTS_Chen,t_DBRTS_Levy} and the Coulomb 
blockade\cite{t_DBRTS_Hung,t_DBRTS_Thielmann} regimes. Noise in QPC
was studied within NEGF\cite{t_QPC_Averin} and kinetic equation 
approaches,\cite{t_QPC_Green} while noise in Coulomb blockaded
Josephson junctions was investigated within a phase correlation theory
approach.\cite{t_JJ_Sonin2} NEGF was also applied to study 
shot noise in chain models\cite{t_chain_Aghassi,t_chain_Kinderman} and
disordered junctions.\cite{t_disordered_Gutman} Shot noise for coherent
electron transport in molecular devices was investigated within
a scattering theory approach in a number of 
works.\cite{t_mol_Dallakyan,t_mol_Walczak,t_mol_DiVentra2}
Inelastic effects in the noise spectrum were studied recently, first in 
connection with nanoelectromechanical systems 
(NEMS)\cite{t_NEMS_Nishiguchi,t_NEMS_Smirnov,t_NEMS_Clerk,t_NEMS_Novotny,
t_NEMS_Flindt,t_NEMS_Wabnig} 
and ac-driven junctions.\cite{t_AC_Camalet,t_AC_Guyon}
Substantial work on this issue has been done within a scattering theory 
approach.\cite{t_stmol_Shimizu,t_stmol_Bo,t_stmol_Dong,t_stmol_DiVentra}
Finally predictions on phonon assisted resonant shot noise spectra
of molecular junctions were obtained in Ref.~\onlinecite{noise_elph_Balatsky} 
within NEGF.

In this work we investigate inelastic tunneling and its 
influence on the zero frequency noise in a simple one-level molecular
junction model. The one-level junction model focuses on one molecular orbital
(say the lowest unoccupied molecular orbital, LUMO) that supports
resonance transmission beyond a certain voltage bias threshold.
The molecular orbital is coupled to a local oscillator representing
the molecular nuclear subsystem. The oscillator in turn is coupled to
a thermal harmonic bath. We have recently used this model within the NEGF 
methodology to study inelastic effects on electronic transport in
molecular junctions in the weak\cite{IETS_SCBA} 
and in the intermediate to strong\cite{strong_el_ph} electron phonon coupling 
regimes. Here we utilize the same approaches
to investigate inelastic effects in the noise spectrum.
The physics of the problem is determined by the relative magnitudes
of the relevant energy parameters: $\Delta E$ - the spacing between the
leads' Fermi energies and the energy of the molecular orbital, $\Gamma$ - the
broadening of the molecular level due to electron transfer interaction
with the leads, $M$ - the electron-phonon coupling and $\omega$ - the
phonon frequency. We consider both resonant and far off-resonant 
inelastic tunneling cases. 
In the off-resonant limit, $\Delta E \gg \Gamma$ it is usually the case
that also $\Delta E \gg M$ and the process can be described in the
weak electron-phonon coupling limit. 
For resonance transmission, $\Delta E < \Gamma$, the electron-phonon 
interaction may still be weak, $M \ll \Gamma$.\cite{weakM} These weak coupling
situations can be treated
within standard diagrammatic perturbation theory on the Keldysh contour,
see e.g.~\onlinecite{IETS_SCBA}. More interesting physics is observed in
the case of strong electron-phonon interaction, $M \gg \Gamma$,
which we treat within the self-consistent NEGF scheme introduced in
Ref.~\onlinecite{strong_el_ph}. In both limits we apply the corresponding
approach to study effects of electron-phonon coupling on the zero-frequency
noise and the Fano factor.

Our results are discussed in comparison with recent work on the effect 
of electron-phonon coupling on the noise properties of molecular junctions.
In the weak coupling limit our result agree qualitatively with the
scattering theory based treatment of Chen and 
Di~Ventra\cite{t_stmol_DiVentra} in the case of near resonance transmission
in symmetric junctions. However, in contrast to 
Ref.~\onlinecite{t_stmol_DiVentra}
that finds that the leading contribution to this effect is of order $M^4$
we find that this effect is of order $M^2$. Also, different qualitative
behavior is found in off-resonance situations and for resonance transmission
through asymmetric junctions.

In the strong electron-phonon coupling limit we compare our result to that
of Zhu and Balatsky\cite{noise_elph_Balatsky} who treat this
problem using the (essentially scattering theory) procedure of 
Lundin and McKenzie.\cite{Lundin} Again we find some new qualitative
effects that can not be obtained from the simpler approach. 

Our analysis suggests the usefulness of the
``noise spectroscopy'', either in a source-drain response or in a gate voltage
response, to measure the values of the electron-vibration coupling
on the bridging molecule and to characterize the asymmetry in the
coupling between the molecule and the source and drain electrodes.

Section~\ref{general} describes formal derivation of the expression
for the noise through a molecular junction.
In Section~\ref{model} we introduce the model.
Section~\ref{elastic} briefly reviews the noise properties
of the elastic current. 
Section~\ref{weak} sketches a procedure and presents numerical 
results for the case of weak electron-phonon coupling
and Section~\ref{strong} does the same for the strong coupling case.
Section~\ref{conclude} concludes.

\section{\label{general}General noise expression}
A general expression for the frequency dependent noise in a 2-terminal 
junction can be obtained by applying the method of Meir and Wingreen.
In its most general formulation this method assumes a single electron exchange
between bridge and leads and invokes the non-crossing approximation 
(NCA)\cite{Bickers} for the bridge-leads coupling. The system is described by
the Hamiltonian
\begin{equation}
\label{genH}
 \hat H = \hat H_M + \hat H_L + \hat H_R + \sum_{k\in\{L,R\}}\sum_{m\in M}
 \left(V_{km}\hat c_k^\dagger\hat c_m + \mbox{H.c.}\right)
\end{equation}
where $\hat H_M$, $\hat H_L$ and $\hat H_R$ denote the Hamiltonian of the
independent molecule and the left and right leads, respectively
and where $\hat c_m$ ($\hat c_m^\dagger$) and $\hat c_k$ ($\hat c_k^\dagger$)
denote single electron annihilation (creation) operators in molecule and 
leads. The current operators for the molecule and each lead are defined by
\begin{equation}
\label{hatIK}
 \hat I_K(t) = \frac{2 i e}{\hbar} \sum_{k\in K}\sum_{m\in M}
 \left(V_{km}\hat c_k^\dagger(t) \hat c_m(t)
              -  V_{mk} \hat c_m^\dagger(t)\hat c_k(t)\right)
\end{equation}
and the average steady state current for a given potential bias is obtained
in the form\cite{current,HaugJauho}
\begin{eqnarray}
\label{IK}
 I_K &=& \left<\hat I_K\right> = \int\frac{dE}{2\pi} i_K(E) \\
\label{iK}
 i_K(E) &=& \frac{2e}{\hbar} 
 \mbox{Tr}\left[\mathbf{\Sigma}_K^{<}(E)\, \mathbf{G}^{>}(E) 
 - \mathbf{\Sigma}_K^{>}(E)\, \mathbf{G}^{<}(E)\right]\end{eqnarray}
where $\mathbf{\Sigma}_K^{<,>}$ are lesser/greater projections of the 
self-energy due to coupling to the (metallic) contact $K$ ($K=L,R$) 
and $\mathbf{G}^{<,>}$ are lesser/greater Green functions, all these 
are defined in a molecular subspace of the problem.
Obviously $I_L=-I_R$ in steady state.

The noise spectrum is defined as the Fourier transform of the current-current
correlation function. The relevant current fluctuations have to be considered
carefully because by their nature such fluctuations correspond to transient
deviations from steady states. We follow the analysis of 
Ref.~\onlinecite{Galperin}
that leads to the following expression for the operator that corresponds to
the current in the outer circuit
\begin{equation}
\label{hatI}
 \hat I(t) = \eta_L\hat I_L(t) + \eta_R \hat I_R(t)
\end{equation}
where
\begin{equation}
\label{eta}
 \eta_L = \frac{C_R}{C} \qquad \eta_R = -\frac{C_L}{C} \qquad C = C_L+C_R
\end{equation}
$C_L$ and $C_R$ are junction capacitance parameters that take into account
the charge accumulation at the corresponding bridge-lead interface.
Note that at steady state $I=I_L=-I_R$. The different signs assigned to
$\eta_L$ and $\eta_R$ stem from the fact that the current is considered
positive on either side when carriers enter the system (molecule). 

In terms of $\hat I$, Eq.~(\ref{hatI}), the noise spectrum is defined by
\begin{eqnarray}
\label{def_noise}
 S(\omega) &=& 2\int_{-\infty}^{+\infty} dt\, S(t) e^{i\omega t} \\
 S(t) &=& \frac{1}{2}\left<\left\{\Delta\hat I(t);\Delta\hat I(0)\right\}\right>\end{eqnarray}
where $\Delta\hat I = \hat I - I$ and where the averaging is done as usual
(and as in Eq.~(\ref{IK})) on the non-interacting state of the system at
the infinite past. The same approach\cite{current,HaugJauho} that leads
to Eq.~(\ref{IK}) now leads to the following general expression for the 
noise
\begin{widetext}
\begin{align}
 \label{S}
 S(\omega) =& \frac{2 e^2}{\hbar} \sum_{K_1,K_2=\{L,R\}} \eta_{K_1}\eta_{K_2}
 \int_{-\infty}^{+\infty}\frac{dE}{2\pi}\mbox{Tr}\left[
  \delta_{K_1,K_2}\left[G^{<}(E-\omega)\,\Sigma_{K_1}^{>}(E)
                        +\Sigma_{K_1}^{<}(E)\,G^{>}(E-\omega)\right]
 \right.\nonumber \\ & 
  + G^{<}(E-\omega)\left[\Sigma_{K_1}\,G\,\Sigma_{K_2}\right]^{>}(E)
       + G^{>}(E-\omega)\left[\Sigma_{K_1}\,G\,\Sigma_{K_2}\right]^{<}(E)
 \\ &
 - \left[\Sigma_{K_1}\,G\right]^{<}(E-\omega)\left[\Sigma_{K_2}\,G\right]^{>}(E)
 \left.
 - \left[G\,\Sigma_{K_1}\right]^{<}(E-\omega)\left[G\,\Sigma_{K_2}\right]^{>}(E) + \left(\omega \to -\omega\right)\right]
 \nonumber
\end{align}
A similar approach was used by Bo and Galperin\cite{noise_Galperin} to
obtain an expression analogous to (\ref{S}) for a single level bridge 
model.\cite{noise_expression} 
In this paper we  restrict our consideration to the case of zero frequency
noise which is the relevant observable when the measurement time is long
relative to the electron transfer time, a common situation in usual
experimental setup. Then the expression simplifies to
\begin{equation}
 \label{Sweq0}
 S(\omega=0) = \frac{4 e^2}{\hbar} \int_{-\infty}^{+\infty}\frac{dE}{2\pi}
 \left[\sum_{K_1,K_2=\{L,R\}} \left( \eta_{K_1}\eta_{K_2}
 \mbox{Tr}
 \left[ \delta_{K_1,K_2}\left[G^{<}\,\Sigma_{K_1}^{>}
                              +\Sigma_{K_1}^{<}\,G^{>}\right]
 - 2\Sigma_{K_1}^{<}\,\Sigma_{K_2}^{>}|G^r|^2 \right]\right)
 - i^2(E)\right] 
\end{equation}
where $i(E)=\sum_{K=\{L,R\}}\eta_K i_K(E)$ with $i_K(E)$ given by 
Eq.~(\ref{iK}).
\end{widetext}

\section{\label{model}The single level bridge/free electron leads model}
Eq.~(\ref{Sweq0}) can be applied, under the approximations specified to very
general bridge and leads models. Further progress can be made by specifying
to simpler models. As usual we assume that the contacts are treated as 
reservoirs of free electrons, each in its own equilibrium. 
The molecule is described by the simplest coupled electron-phonon system 
as follows.
Assuming that the energy gap between the orbitals relevant for a molecular 
junction (in particular the HOMO-LUMO gap) is much greater than level
broadening due to coupling to the contacts, we consider the electron current
through each orbital separately. Thus the molecular junction is represented by
a single level (molecular orbital) coupled to two contacts ($L$ and $R$). 
The electrons on the bridge are coupled to a local normal mode 
(henceforth referred to as primary phonon) represented by a harmonic oscillator.
This is coupled to a thermal bath described by a set of independent harmonic 
oscillators (``secondary phonons'').  The Hamiltonian is thus given by
\begin{align}
 \label{H}
 &\hat H = \varepsilon_0\hat c^\dagger\hat c +
 \sum_{k\in\{L,R\}} \varepsilon_k \hat c_k^\dagger\hat c_k +
 \sum_{k\in\{L,R\}} \left(V_k\hat c_k^\dagger\hat c + \mbox{h.c.}\right)
 \nonumber \\ &+
 \hbar\omega_0\hat a^\dagger\hat a +
 \sum_\beta\hbar\omega_\beta\hat b^\dagger_\beta\hat b_\beta +
 M_a\hat Q_a\hat c^\dagger\hat c +
 \sum_\beta U_\beta\hat Q_a\hat Q_\beta
\end{align}
where $\hat c^\dagger$ ($\hat c$) are creation (destruction) operators
of electrons on the level, $\hat c_k^\dagger$ ($\hat c_k$) are the corresponding
operators for electronic states in the contacts, $\hat a^\dagger$ ($\hat a$)
are creation (destruction) operators for the primary phonon, and
$\hat b_\beta^\dagger$ ($\hat b_\beta$) are corresponding operators for
phonon states in thermal (phonon) bath.
$\hat Q_a$ and $\hat Q_\beta$ are shift operators
\begin{equation}
 \label{Q}
 \hat Q_a = \hat a + \hat a^\dagger \qquad
 \hat Q_\beta = \hat b_\beta + \hat b_\beta^\dagger
\end{equation}
Detailed discussion of the model and our self-consistent calculation procedures
are presented in Refs.~\onlinecite{IETS_SCBA} and \onlinecite{strong_el_ph} for
weak and strong electron-phonon couplings respectively. Sections
\ref{weak} and \ref{strong} below provide brief outlines of these procedures. 

In both weak and strong coupling cases the calculation is aimed to
obtain the electron Green function (or rather its Langreth 
projections\cite{Langreth} on the real time axis). Those Green functions are
used for the calculation of the steady-state current across the 
junction (\ref{IK}) and the noise spectrum (\ref{Sweq0}).
The general current and noise expression remain as before, Eqs.~(\ref{IK}), 
(\ref{S}) and (\ref{Sweq0}) (excluding the trace operations).
The lesser and greater self-energies associated with the bridge-contact
coupling are now given by
\begin{eqnarray}
 \label{SEKlt}
 \Sigma_K^{<}(E) &=& i f_K(E) \Gamma_K(E) \\
 \label{SEKgt}
 \Sigma_K^{>}(E) &=& -i [1-f_K(E)] \Gamma_K(E)
\end{eqnarray}
with $f_K(E)$ the Fermi distribution in the contact $K=L,R$ and
\begin{equation}
 \label{GammaK}
 \Gamma_K(E) = 2\pi \sum_{k\in K} |V_k|^2 \delta(E-\varepsilon_k)
\end{equation}
In the calculation described below we adopt the wide band 
approximation in which $\Gamma_K$ is assumed constant. 
In this approximation $\Sigma^r_K$=$\left[\Sigma^a_K\right]^{*}=-i\Gamma_K/2$.

The observed conduction and noise properties of our system depend on
parameters that can not be accounted for by the present single-electron
level of treatment. The capacitances $C_L$ and $C_R$ that define the 
parameters $\eta_L$ and $\eta_R$ that enter the noise calculation 
belong to this group. We also use below the voltage division parameter $\delta$
that determines the voltage induced shifts in the leads' electrochemical 
potentials $\mu_L$ and $\mu_R$ relative to $\varepsilon_0$ according to
\begin{equation}
\label{delta}
 \begin{array}{rclcl}
 \mu_L&=&E_F+\delta\, eV &\equiv& E_F+eV_L  \\
 \mu_R&=&E_F-(1-\delta)\, eV &\equiv& E_F+eV_R
 \end{array}
\end{equation}
This parameter affects the way by which the externally imposed bias
translates into the relative positioning of electronic energies.

In addition, consider the ratio
\begin{equation}
\label{alpha}
 \alpha = \frac{\Gamma_L}{\Gamma} \qquad 1-\alpha = \frac{\Gamma_R}{\Gamma}
\end{equation}
that represents the asymmetry in molecule-lead couplings (wide band limit
is assumed here and below). This is an experimentally controlled parameter
that can be changed, e.g. with tip-molecule distance in an STM configuration
or by using thin insulating layers to separate molecule from 
leads.\cite{Ho,Repp}

The connection (if any) between the parameters $\alpha$, $\delta$ and 
$\eta\equiv\eta_L$ ($\eta_R=\eta_L-1$) is obviously of great interest.
An equivalent circuit model for a molecular (or nanodot) 
junction\cite{Galperin,Tinkham,IngoldNazarov} usually describes the
junction as serially connected $RC$ circuits. Assuming that at steady state 
charges $q_L$ and $q_R$ accumulate on the corresponding capacitors the
steady state relationships $V_L=q_L/C_L=IR_L=\Gamma_Lq_LR_L$ imply that
$\Gamma_L=(R_LC_L)^{-1}$. These relations imply in turn that
\begin{equation}
\label{params}
 \frac{\alpha}{1-\alpha} = \frac{1-\delta}{\delta}\cdot\frac{1-\eta}{\eta}
\end{equation}
This leaves two important parameters in the model. These however can,
at least in principle be determined experimentally: The rates $\Gamma$
can be determined by optical experiments\cite{Silbey} and the potential
distribution on a conducting junction can be probed as demonstrated in
Ref.~\onlinecite{McEuen}. In the present paper we use 
$\alpha=\Gamma_L/\Gamma$ to characterize the junction 
asymmetry and several arbitrary values of $\eta=\eta_L=C_L/C$
chosen to represent different possible manifestations of noise
behavior.

In addition, in this paper we discuss two experimental manifestations 
of nanojunction
transport. In the first, the source-drain voltage $V=(\mu_L-\mu_R)/e$
is kept small and fixed, $k_B T\sim eV\ll\Gamma$, and the gate voltage
is varied in order to move the molecular orbital energy in and out of
resonance. In the second, the gate voltage $V_g$ is fixed, e.g. $V_g=0$,
and $V$ varies so that resonance conditions are usually achieved for
$k_B T\ll\Gamma\ll eV$.  In the latter case 
we restrict our consideration to the case where the onset of resonance 
transmission is caused by $\mu_L$ crossing the LUMO for a positively biased 
junction, and disregard other possible scenarios such as the  $\mu_R$ crossing 
the HOMO.\cite{muRHOMO} 
In both cases taking $k_BT\ll\Gamma$ provides better resolution of
resonance features.
Also, in the resonant tunneling situation it is convenient to consider 
the first derivatives of the current and noise with respect to voltage,
$dI/dV$ and $dS(\omega=0)/dV$ respectively, rather than $I$ and 
$S(\omega=0)$ themselves.

\section{\label{elastic}Noise in elastic transport}
Before discussing inelastic effects we briefly review the
issue of noise in the elastic tunneling situation. 
In this case current, Eq.(\ref{IK}), reduces to
the well known Landauer expression
\begin{equation}
 \label{I0}
 I = \frac{e}{\hbar}\int_{-\infty}^{+\infty}\frac{dE}{2\pi}\,
 T_0(E)\left[f_L(E)-f_R(E)\right]
\end{equation}
while the zero frequency noise expression (\ref{Sweq0}) reduces to 
a sum of thermal contribution, $S_t=S_t(\omega=0)$, due to thermal excitations 
in the contacts and shot noise term, $S_s=S_s(\omega=0)$, due to the discrete 
nature of the electron transport\cite{noiserev_Buttiker}
\begin{eqnarray}
 \label{S0}
 S &=& S(\omega=0) = S_t(\omega=0) + S_s(\omega=0) \\
 \label{S0t}
 S_t &=& \frac{4e^2}{\hbar}\int_{-\infty}^{+\infty}\frac{dE}{2\pi}\,
            T_0(E)
 \\ &\times&
 \left\{f_L(E)[1-f_L(E)]+f_R(E)[1-f_R(E)]\right\} 
 \nonumber\\
 \label{S0s}
 S_s &=& \frac{4e^2}{\hbar}\int_{-\infty}^{+\infty}\frac{dE}{2\pi}\,
            T_0(E)\left\{1-T_0(E)\right\}
 \\ &\times&
 \left[f_L(E)-f_R(E)\right]^2
 \nonumber
\end{eqnarray}
with
\begin{equation}
 \label{T0}
 T_0(E) = \frac{\Gamma_L\Gamma_R}{(E-\varepsilon_0)^2+(\Gamma/2)^2}
\end{equation}
Using Eq.~(\ref{delta}) this leads to (with $\beta^{-1}=k_BT$)
\begin{widetext}
\begin{align}
\label{dSt_dV}
 &\frac{\partial S_t}{\partial V} = \frac{4e^3\beta}{\hbar}
 \int_{-\infty}^{+\infty}\frac{dE}{2\pi}\, T_0(E)
 \left[\delta\frac{\sinh(\beta(E-\mu_L)/2)}{\cosh^3(\beta(E-\mu_L)/2)}
     -(1-\delta)\frac{\sinh(\beta(E-\mu_R)/2)}{\cosh^3(\beta(E-\mu_R)/2)}\right]
 \\
\label{dSs_dV}
 &\frac{\partial S_s}{\partial V} = \frac{4e^3\beta}{\hbar}
 \int_{-\infty}^{+\infty}\frac{dE}{2\pi}\, T_0(E)[1-T_0(E)]\,
 2[f_L(E)-f_R(E)]
 \left[\frac{\delta}{4\cosh^2(\beta(E-\mu_L)/2)}
      +\frac{1-\delta}{4\cosh^2(\beta(E-\mu_R)/2)}\right]
\end{align}
\end{widetext}
Interestingly, these results do not depend on the capacitance ratios $\eta$.
These expressions can be further simplified in two cases: 
(a) When the source-drain voltage $V$ is varied we consider, 
as indicated above, the situation where the signal is dominated by 
the resonance associated with the crossing
between $\varepsilon_0$ and $\mu_L$. In this case the terms involving $\mu_R$
in Eqs.~(\ref{dSt_dV}) and (\ref{dSs_dV}) are disregarded.
(b) When the gate voltage $V_g$ is varied at small $V$ the two terms on the
r.h.s. of Eqs.~(\ref{dSt_dV}) and (\ref{dSs_dV}) are essentially the same
and can be combined together.  

In these elastic transmission cases both experimental procedures 
discussed above give essentially the same information. 
Recall that conductance $dI/dV$ is a Lorentzian 
function of $V$ that peaks at the position of the resonance level, 
with a width (FWHM) related to the total escape rate $\Gamma$. 
In contrast, the differential noise graph ($dS/dV$ vs. $V$) depends
on the asymmetry of the molecule coupling in the
junction $\alpha$. 
$S_t$ itself is a Lorentzian function of $V$, so $dS_t/dV$ vs. $V$ has 
the form of a Lorentzian derivative that vanishes at the level position
and peaks at $\varepsilon_0\mp \Gamma/2$ with peak/dip heights 
$\sim\pm \frac{\Gamma_L\Gamma_R}{\Gamma^2}\frac{k_B T}{\Gamma}$
respectively.\cite{heightasym} 
The contribution from the zero temperature shot noise, 
$dS_s/dV$ vs. $V$, depends on the coupling asymmetry $\alpha$.
A double peak structure appears when $T_0(E)[1-T_0(E)]$ has three
extrema as a function of $E$. The condition for this to occur is
\begin{equation}
 \label{cond_dp_0}
 \alpha^2 - \alpha + \frac{1}{8} < 0
\end{equation}
so that for $\frac{1-1/\sqrt{2}}{2}<\alpha<\frac{1+1/\sqrt{2}}{2}$
it yields a two peak structure with peaks at 
$E=\varepsilon_0\mp[\Gamma/2]\sqrt{8\alpha(1-\alpha)-1}$
and peak heights of $1/4$ (in units of $4e^3/\hbar$).
This behavior of $S_s(V)$ is associated with the transition from 
double-barrier resonance structure at moderate asymmetry to a single 
tunneling barrier for the strongly asymmetric case.
$S_s$ vanishes when $T_0=0$ or $T_0=1$ and maximizes when $T_0=1/2$. 
This leads to a two peak structure in symmetric and nearly symmetric 
double barrier resonance supporting tunnel junctions (DBRTS).
When the asymmetry is stronger the noise profile has a single peak form with 
peak position at $\varepsilon_0$. 
$dS_s/dV|_{\varepsilon_0}$ in both cases
is given by $16\alpha(1-\alpha)[1/4-\alpha(1-\alpha)]$ 
(in units of $4e^3/\hbar$), which is $0$ for
symmetric coupling case ($\alpha=1/2$). 

\begin{figure}[t]
\centering\includegraphics[width=\linewidth]{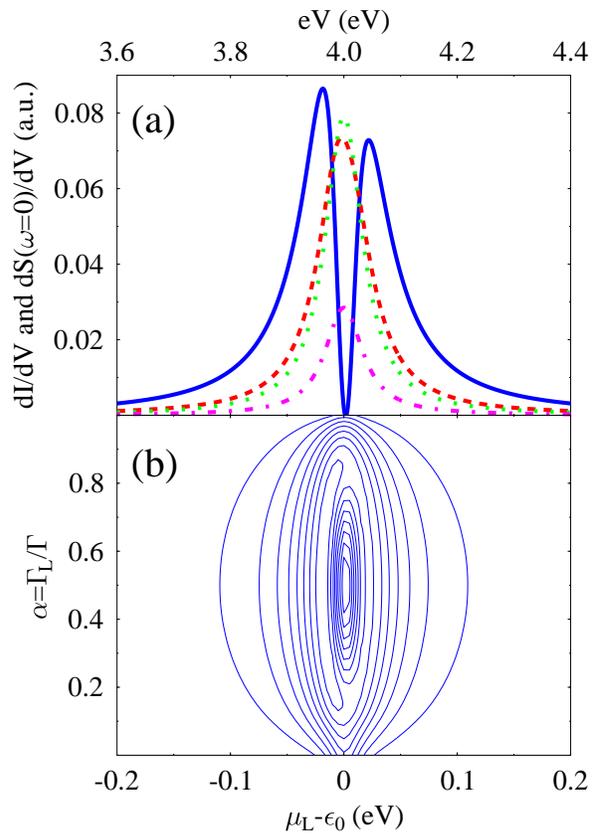}
\caption{\label{fig_noise0_a}(Color online) 
Conductance and differential noise vs. applied
source-drain voltage. (a) $dS(\omega=0)/dV$ (solid and dashed lines) and
$dI/dV$ (dotted and dashed-dotted lines) for $\alpha=0.5$ and $\alpha=0.1$
respectively. (b) Contour plot of $dS(\omega=0)/dV$. See text for units.
}
\end{figure}

The sum $S_t+S_s$ determines the observed signal. 
Figure~\ref{fig_noise0_a} illustrates the above
discussion using the parameters $\varepsilon_0=2$~eV, $\Gamma=0.04$~eV, 
and $T=10$~K. 
Here and below we use a.u. (atomic units) for current, noise and 
differential noise, $e/t_0$, $e^2/t_0$, and $e^3/\hbar$ respectively,
with $e$ the electron charge, $t_0$ the atomic unit of time, 
and $\hbar=h/2\pi$ the Planck constant.
Fig.~\ref{fig_noise0_a}a shows plots of the conductance and the
differential noise for symmetric ($\alpha=0.5$) and strongly
asymmetric ($\alpha=0.1$) coupling. In both cases $dI/dV$ 
is a Lorentzian (dotted and dash-dotted lines respectively), while
form of $dS/dV$ changes from a two-peak structure with complete noise 
suppression at $\mu_L=\varepsilon_0$ for the symmetric case to a one peak
structure in the asymmetric situation. Note  that in the two-peak case
the peak heights are different due to the thermal noise contribution.
In principle both the peaks positions in the two-peak structure and 
the differential noise at $\mu_L=\varepsilon_0$ 
may be used as sources of information on the asymmetry (expressed by the
parameter $\alpha$) in the molecule-leads couplings.
A single peak $dS/dV$ shape indicates highly asymmetric coupling. 
Consequently, in STM experiments the shape of the
differential noise vs. applied voltage plot will change when bringing
the tip closer to the sample, decreasing the $\Gamma_L/\Gamma_R$
asymmetry.

\begin{figure}[t]
\centering\includegraphics[width=\linewidth]{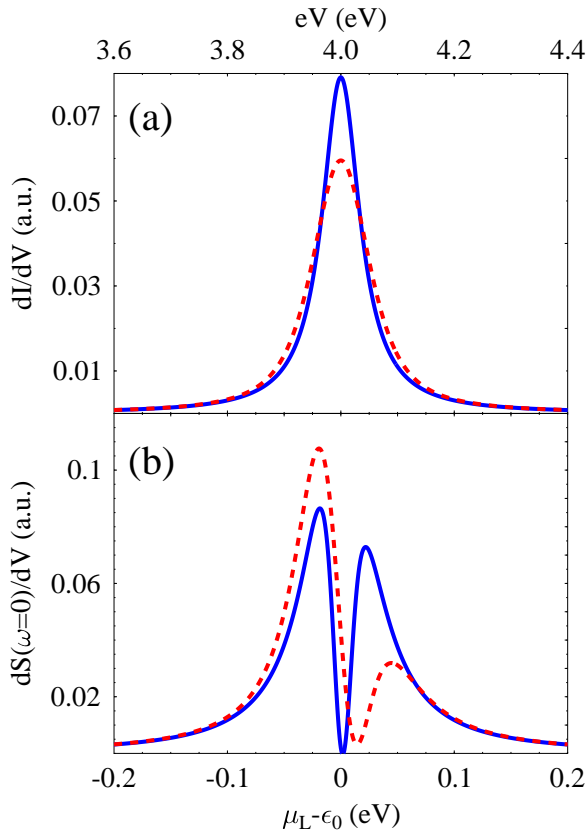}
\caption{\label{fig_noise0_t}(Color online) 
Conductance (a) and differential noise
(b) vs. applied source-drain voltage. Shown are results for $T=10$~K
(solid line) and $T=100$~K (dashed line).
}
\end{figure}

The effect of temperature is demonstrated in Fig.~\ref{fig_noise0_t} where
results for two temperatures, $T=10$~K and $T=100$~K respectively, 
are presented for the symmetric coupling case, $\alpha=0.5$.
The differential conductance peak, as is expected, becomes wider and lower
with increasing $T$ (Fig.~\ref{fig_noise0_t}a). 
At the same time the two peak structure of the differential noise plot
becomes broader, and the difference in the peak 
heights of the two-peak structure in the differential noise plot 
becomes more pronounced (Fig.~\ref{fig_noise0_t}b).

\section{\label{weak}Weak electron-phonon coupling}
We now turn to the model, Eq.~(\ref{H}), that includes the effect of
electron-phonon interactions on the bridge. In this section
we consider weak electron-phonon coupling which can be treated
with the standard diagrammatic theory on the Keldysh contour.
The lowest non-zero diagrams lead to the Born approximation (BA), while
dressing of these diagrams yields its self-consistent version (SCBA).
Detailed description of this approach and the derivation of 
the self-energies expressions can be found in our previous 
publication.\cite{IETS_SCBA}
Within the Born approximation the lesser and greater electron self-energies 
due to coupling to the local phonon can be expressed for the model at hand 
in terms of the (molecule projected) electron and phonon densities of states, 
$\rho_{el}(E)=-\mbox{Im}[G^r(E)]/\pi$ and 
$\rho_{ph}(\omega)=-\mbox{sgn}(\omega)\mbox{Im}[D^r(E)]/\pi$ 
respectively, and of the electron and phonon energy-resolved occupations, 
$n(E)$ and $N(\omega)$ respectively,\cite{occ} as follows 
(for details see Ref.~\onlinecite{IETS_SCBA})
\begin{align}
 \label{Sphgt}
 &\Sigma_{ph}^{>}(E) = -2\pi iM_a^2 \int_0^\infty d\omega\,
 \rho_{ph}(\omega)
 \nonumber \\ &\times
 \left\{ 
 [1+N(\omega)]\rho_{el}(E-\omega)[1-n(E-\omega)]
 \right.\\ &\quad\left.
 +N(\omega)\rho_{el}(E+\omega)[1-n(E+\omega)]
 \right\}
 \nonumber
\end{align}
\begin{align}
 \label{Sphlt}
 &\Sigma_{ph}^{<}(E) = 2\pi iM_a^2 \int_0^\infty d\omega\,
 \rho_{ph}(\omega)
 \nonumber\\ &\times
 \left\{
 N(\omega)\rho_{el}(E-\omega)n(E-\omega)
 \right.\\ &\quad\left.
 +[1+N(\omega)]\rho_{el}(E+\omega)n(E+\omega)
 \right\}
 \nonumber
\end{align}
Note that in the 1-level bridge model employed here all the Green
functions and self-energies are scalars defined in the bridge subspace.
In the absence of electron-phonon coupling the densities of states are
\begin{eqnarray}
 \label{rhoel}
 \rho_{el}(E) &=& \frac{\Gamma/2\pi}{(E-\varepsilon_0)^2+(\Gamma/2)^2}
 \\
 \label{rhoph}
 \rho_{ph}(\omega) &=& 
 \frac{\gamma_{ph}/2\pi}{(\omega-\omega_0)^2+(\gamma_{ph}/2)^2}
\end{eqnarray}
where $\Gamma=\Gamma_L+\Gamma_R$, see Eq.~(\ref{GammaK}),
is resonant level escape rate due to 
coupling to the leads and $\gamma_{ph}$ is the local phonon decay rate
due to coupling to the thermal bath. The phonon occupation $N(\omega)$ in 
this limit is given by the Bose-Einstein distribution, 
$N(\omega)=\left[e^{\beta\hbar\omega}-1\right]^{-1}$,
while the electron occupation in the junction is
\begin{equation}
 \label{nel}
 n(E) = \frac{\Gamma_L}{\Gamma}f_L(E) + \frac{\Gamma_R}{\Gamma}f_R(E)
\end{equation}
where $f_K(E)$ is Fermi-Dirac distribution of the contact $K$. 
The electron relaxation rate due to coupling to the primary phonon is 
given by
\begin{equation}
 \label{Gamph}
 \Gamma_{ph}(E) = i[\Sigma_{ph}^{>}(E)-\Sigma_{ph}^{<}(E)]
\end{equation}
The retarded electron self-energy due to this coupling
is taken in what follows to be purely imaginary (tacitly assuming that
its real part -- level shift -- has been incorporated into
$\varepsilon_0$), given by
\begin{equation}
 \label{Sphr}
 \Sigma_{ph}^{r}(E) = -i\frac{\Gamma_{ph}(E)}{2} 
\end{equation}
Keeping terms up to the second order in the electron-phonon 
coupling we get for the electron Green functions
\begin{align}
 \label{Gr}
 &G^r(E) = G_0^r(E) -i G_0^r(E) \frac{\Gamma_{ph}(E)}{2} G_0^r(E)
 \\
 \label{Gltgt}
 &G^{<,>}(E) = \left|G_0^r(E)\right|^2
 \left\{
 \left[1-\pi\Gamma_{ph}(E)\rho_{el}(E)\right]
 \right. \\ &\left.\qquad\times
 \left[\Sigma_L^{<,>}(E)+\Sigma_R^{<,>}(E)\right]
 + \Sigma_{ph}^{<,>}(E) \right\}
 \nonumber
\end{align}
where
\begin{equation}
 \label{G0r}
 G_0^r(E) = \left[E-\varepsilon_0+i\Gamma/2\right]^{-1}
\end{equation}
and where the lesser and greater electron self-energies due to coupling 
to contacts, $\Sigma_K^{<,>}(E)$, $K=L,R$ are given by 
Eqs.~(\ref{SEKlt}) and (\ref{SEKgt}).

Using (\ref{SEKlt}), (\ref{SEKgt}), and (\ref{Gltgt}) in (\ref{IK}) 
the average steady-state current through the junction is obtained in the form
\begin{equation}
 \label{current_weakelph}
 I = \frac{2e}{\hbar}\int_{-\infty}^{+\infty}\frac{dE}{2\pi}\,
     T(E)\left[f_L(E)-f_R(E)\right]
\end{equation}
where 
\begin{equation}
 \label{T}
 T(E) = T_0(E)\left[1+\frac{(E-\varepsilon_0)^2-(\Gamma/2)^2}
                           {(E-\varepsilon_0)^2+(\Gamma/2)^2}
        \frac{\Gamma_{ph}(E)}{\Gamma}\right]
\end{equation}
with $T_0(E)$ given in (\ref{T0}).
The first term in parentheses in (\ref{T}) yields usual Landauer expression,
while the second gives the second order (in $M_a$) correction 
associated with the electron-phonon coupling. As discussed in 
Ref.~\onlinecite{IETS_SCBA}, this correction results
from both the renormalization of the elastic transmission channel 
and the opening of the inelastic channel.

Using (\ref{SEKlt}), (\ref{SEKgt}), (\ref{Gr}), (\ref{Gltgt}) in (\ref{Sweq0})
we get for the zero frequency noise
\begin{widetext}
\begin{align}
 \label{Sweq0_weakelph}
 &S(\omega=0) = \frac{4e^2}{\hbar}\int_{-\infty}^{+\infty}\frac{dE}{2\pi}\,
 \left\{ T_0(E)B_1(E)
 \left( f_L(E)[1-f_L(E)] + f_R(E)[1-f_R(E)] \right) \right.
 \nonumber \\ &
 +T_0(E)B_1(E) \left(1-T_0(E)B_1(E)\right) \left[f_L(E)-f_R(E)\right]^2
 \\ &\left.
 -2\frac{T_0^2(E)B_1(E)}{\Gamma_L\Gamma_R}
  \left( f_L(E)-f_R(E) \right) B_2(E)
 - \frac{T_0^2(E)}{(\Gamma_L\Gamma_R)^2} B_2^2(E)
 + \frac{T_0(E)}{\Gamma_L\Gamma_R} B_3(E)
 \right\}
 \nonumber
\end{align}
where
\begin{eqnarray}
 B_1(E) &=& \left[1-\pi\Gamma_{ph}(E)\rho_{el}(E)\right]
 \\
 B_2(E) &=& 
 i\Sigma_{ph}^{>}(E)\left(\eta_L\Gamma_Lf_L(E)+\eta_R\Gamma_Rf_R(E)\right) 
 +
 i\Sigma_{ph}^{<}(E)\left(\eta_L\Gamma_L[1-f_L(E)]+\eta_R\Gamma_R[1-f_R(E)]\right)
 \\
 B_3(E) &=& 
 i\Sigma_{ph}^{>}(E)\left(\eta_L^2\Gamma_Lf_L(E)+\eta_R^2\Gamma_Rf_R(E)\right) 
 -
 i\Sigma_{ph}^{<}(E)\left(\eta_L^2\Gamma_L[1-f_L(E)]+\eta_R^2\Gamma_R[1-f_R(E)]\right)
\end{eqnarray}
\end{widetext}
One sees that in the presence of electron-phonon interaction 
the noise expression 
can not be separated into thermal and shot noise parts as is given by
(\ref{S0})-(\ref{S0s}).

Eqs.~(\ref{current_weakelph}) and (\ref{Sweq0_weakelph}) are used 
in the calculations presented below. 
We will consider two limits of the electron tunneling process: 
1.~resonant, $\Gamma\gg|E-\varepsilon_0|$, 
and 2.~off-resonant (superexchange), $\Gamma\ll|E-\varepsilon_0|$, 
where $E$ is the energy of the tunneling electron. The first situation was
considered in Ref.~\onlinecite{t_stmol_DiVentra}. The second is
a common situation in molecular junctions where the HOMO-LUMO gap 
is larger than both the orbital broadening and applied voltage. 
It was pointed out by us and others that since 
$\frac{(E-\varepsilon_0)^2-(\Gamma/2)^2}{(E-\varepsilon_0)^2+(\Gamma/2)^2}$
is negative (positive) in the resonant (non-resonant) situations
the current change upon the onset of inelastic tunneling, $\delta I$, 
is negative in the resonant case and positive in the off-resonance limit.

\begin{figure}[t]
\centering\includegraphics[width=\linewidth]{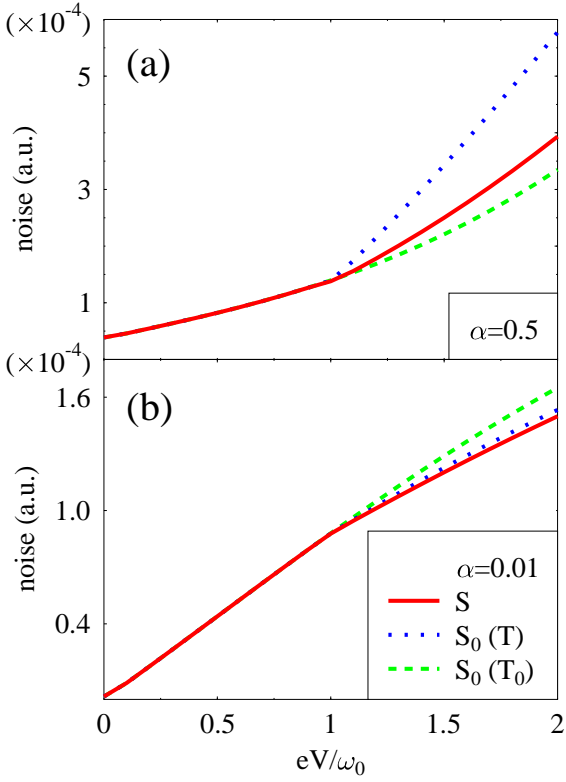}
\caption{\label{fig_S0_Ttot}(Color online)
Zero frequency noise with (solid line) and without (dashed line)
electron-phonon interaction for (a) symmetric, $\alpha=0.5$, and (b) asymmetric,$\alpha=0.01$ molecule-leads coupling. Also shown is approximate
result (dotted line) where noise is calculated as a sum of thermal and shot
noise, Eqs.~(\ref{S0})-(\ref{S0s}),
with renormalized transmission coefficient T, Eq.~(\ref{T}),
replacing the elastic tunneling transmission $T_0$, Eq.~(\ref{T0}).
See text for parameters.}
\end{figure}

Figure~\ref{fig_S0_Ttot} presents the zero frequency noise $S(\omega=0)$,
displayed against the applied voltage for the resonant tunneling case in 
(a) symmetric coupling, $\alpha=0.5$, 
and (b) strongly asymmetric coupling, $\alpha=0.01$, situations. 
Parameters of the calculation are $T=10$~K, $\varepsilon_0=0.05$~eV, 
$\Gamma=0.5$~eV, $\eta=0.5$, $E_F=0$~eV, $\omega_0=0.1$~eV, 
$\gamma_{ph}=0.01$~eV, and $M_a=0.1$~eV. 
The full and dashed lines in Fig.~\ref{fig_S0_Ttot} correspond to
the calculated noise in the presence and absence of electron-phonon
interaction, respectively. The dotted line results from an attempted
approximation in which the noise was computed from Eqs.~(\ref{S0})-(\ref{S0s})
except that $T_0(E)$ was replaced by $T(E)$, Eq.~(\ref{T}).
Two points are noteworthy here. First,
opening of the inelastic channel, at $V/\omega_0=1$, may lead both to 
increase (symmetric coupling, Fig.~\ref{fig_S0_Ttot}a) and decrease
(asymmetric coupling, Fig.~\ref{fig_S0_Ttot}b) of the noise signal.
Second, modifying Eqs.~(\ref{S0})-(\ref{S0s}) by replacing the bare
transmission $T_0(E)$ by the phonon-renormalized transmission $T(E)$
shows a similar qualitative dependence of the noise on the electron-phonon
interaction as our more rigorous results, but fails quantitatively
(compare full and dotted lines in Fig.~\ref{fig_S0_Ttot}a).  

While the calculations presented below are made with no additional assumptions,
a valuable insight can be gained by considering the low temperature limit, 
$T\to 0$, assuming $\omega_0\ll\Delta E$ and/or $\omega_0\ll\Gamma$,
and taking the voltage large enough so that the inelastic channel 
is open, $\mu_L-\mu_R=|eV|>\hbar\omega_0$.
In this case $[f_L(E)-f_R(E)]^2\approx f_L(E)[1-f_R(E)]\approx |f_L(E)-f_R(E)|$
and $f_K(E)[1-f_K(E\pm\omega_0)]\approx 0$.
Under these additional assumptions we obtain for
the noise change upon the onset of inelastic tunneling
(up to $M_a^2$) from (\ref{Sweq0_weakelph}) in the resonant tunneling situation
\begin{align}
 \label{dS_DiVentra}
 &\delta S(\omega=0) = \frac{4e^2}{\hbar}
 \int_{-\infty}^{+\infty}\frac{dE}{2\pi}\, T_0(E)\frac{\Gamma_{ph}(E)}{\Gamma}
 \times\\
 & [f_L(E)-f_R(E)]
 \left\{2(T_0(E)-1)+\eta_L^2+\eta_R^2]
 -2\frac{\Gamma_L\Gamma_R}{\Gamma^2}\right\}
 \nonumber \\
 \label{T0res}
 &T_0(E) \approx \frac{4\Gamma_L\Gamma_R}{\Gamma^2}
\end{align}
Using (\ref{T0res}) in (\ref{dS_DiVentra}) 
one can show that for symmetric coupling, $\alpha=\Gamma_L/\Gamma=0.5$,
$\delta S>0$. This together with $\delta I<0$ in the resonant tunneling case
leads to (predicted in Ref.~\onlinecite{t_stmol_DiVentra}) 
increase in Fano factor, $F=S/I$, upon opening of the inelastic channel. 
The situation can be different however for asymmetric junctions.
Indeed, Eqs.~(\ref{dS_DiVentra}) and (\ref{T0res}) imply that if the
inequality
\begin{equation}
 \alpha^2-\alpha+\frac{2-\eta_L^2-\eta_R^2}{8}>0
\end{equation}   
is satisfied the change in the zero frequency shot noise is negative, 
$\delta S<0$. In this case the Fano factor can decrease with the opening 
of inelastic channel.

\begin{figure}[t]
\centering\includegraphics[width=\linewidth]{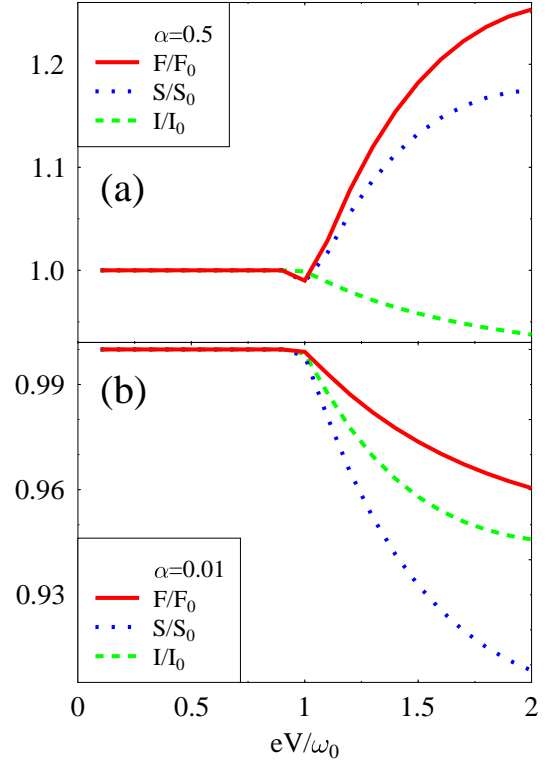}
\caption{\label{fig_m2_diventra}(Color online)
Ratios of Fano factors (solid line), zero frequency noises (dotted
line), and currents (dashed line) with and without electron-phonon coupling
vs. applied voltage in the resonant tunneling regime.
Shown are (a)~symmetric $\alpha=0.5$ and (b)~asymmetric $\alpha=0.01$ coupling
case. See text for parameters.}
\end{figure}

Figures~\ref{fig_m2_diventra}a and \ref{fig_m2_diventra}b (which were obtained,
as already noted, without invoking these simplifying assumptions)
validate these observations. Fig.~\ref{fig_m2_diventra}a shows, for a symmetric
junction a trend (smaller current, larger noise and consequently larger
Fano factor in the presence of electron-phonon interaction) similar to that
suggested by Chen and Di~Ventra\cite{t_stmol_DiVentra} (see Fig.~2 there).
However the trend shown in Fig.~\ref{fig_m2_diventra}b for the asymmetric 
junction is opposite. It should also be noticed that
while the phonon induced
change in the noise was obtained in \onlinecite{t_stmol_DiVentra} to be of order
$\sim M_a^4$, Eq.~(\ref{dS_DiVentra}) shows contributions of order
$\sim M_a^2$ that dominate in the weak electron-phonon coupling limit.

\begin{figure}[htpb]
\centering\includegraphics[width=\linewidth]{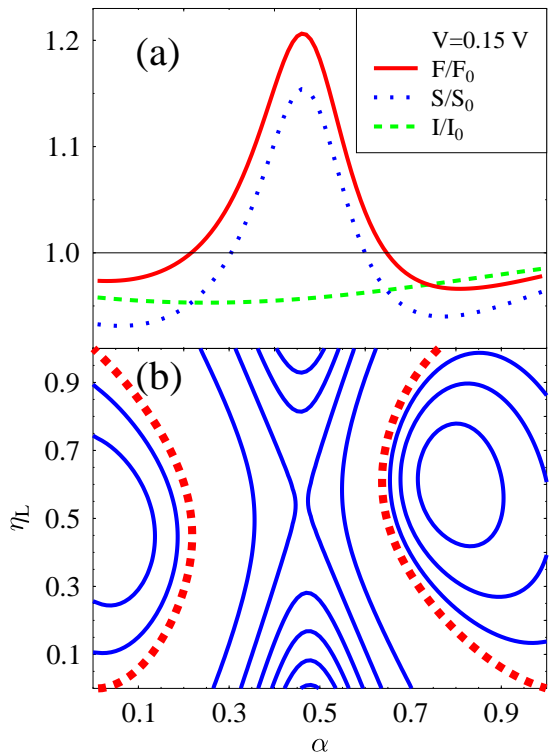}
\caption{\label{fig_m2_a}(Color online)
(a) Ratios of Fano factors ($F/F_0$, solid line), 
zero frequency noises ($S/S_0$, dotted line), 
and currents ($I/I_0$, dashed line) with and without electron-phonon coupling
vs. asymmetry in coupling to the leads in the resonant tunneling regime.
(b) $F/F_0$ a function of two asymmetry parameters,
$\alpha$ and $\eta_L$ in the resonant tunneling regime. Dotted line represents
$F/F_0=1$ case.
Parameters are the same as in Fig.~\ref{fig_m2_diventra} and the voltage
bias is $V=0.15$~V ($eV=1.5\omega_0$).}
\end{figure}

Figure~\ref{fig_m2_a}a shows change in the noise properties with changing 
asymmetry in coupling to the leads as expressed by the parameter $\alpha$.
Note that the current and noise are slightly asymmetric about the symmetric
($\alpha=0.5$) point.
This asymmetry can be rationalized by noting that the difference 
$\Gamma_L\neq\Gamma_R$ implies different tunneling distances 
between the molecule and the left and right leads, resulting in different 
tunneling probabilities towards these leads for an electron that lost
energy to molecular vibrations.
Figure~\ref{fig_m2_a}b gives a broader view of this issue.
It shows a two-dimensional plot of the ratio of Fano factor with and without
electron-phonon coupling,  $F/F_0$, as a function of 
two asymmetry parameters, $\alpha$ and $\eta_L$. At and about the symmetric
point $\alpha=0.5$ this ratio is greater than 1, but it becomes smaller
(and can be smaller than 1) in strongly asymmetric situations, as shown.

The qualitative observations of Ref.~\onlinecite{t_stmol_DiVentra} are seen to
hold only for symmetric molecule-lead coupling but not for highly asymmetric 
situations as encountered in STM configurations.

\begin{figure}[htpb]
\centering\includegraphics[width=1.2\linewidth]{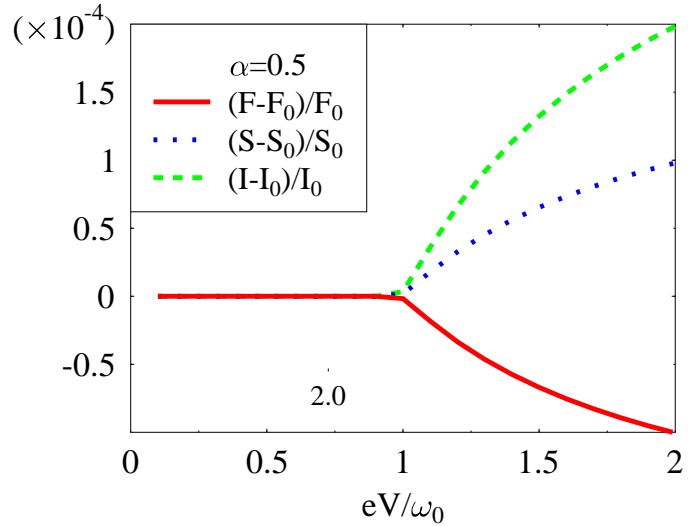}
\caption{\label{fig_m2_antidiventra}(Color online)
Ratios of Fano factors (solid line), zero frequency noises (dotted
line), and currents (dashed line) with and without electron-phonon coupling
vs. applied voltage in the off-resonant tunneling regime.
See text for parameters.}
\end{figure}

Consider next the off-resonant limit, $\Gamma\ll |E-\varepsilon_0|$.
In this case $\delta I$ is positive. When also $\omega_0\ll|E-\varepsilon_0|$
and $T\to 0$ we can use the argument above Eq.~(\ref{dS_DiVentra})
to show that $\delta S>0$. 
In the absence of electron-phonon interaction $I_0\sim T_0\, eV$ and 
$S_0\sim T_0[1-T_0]\, eV$ one gets $F_0=S_0/I_0\sim 1-T_0$. 
When this interaction is present $I\sim T_0[1+\Gamma_{ph}/\Gamma]\, eV$
and $S\sim\{T_0[1-T_0]+T_0\Gamma_{ph}/\Gamma[\eta_L^2+\eta_R^2-2T_0]\}\, eV$, 
so that
\begin{equation}
 \frac{F}{F_0} \sim 1-\frac{\Gamma_{ph}}{\Gamma}
 \left[1-\eta_L^2-\eta_R^2\right] < 1
\end{equation}
Figure~\ref{fig_m2_antidiventra} (which again does not rely on the extreme
limit used to develop our intuition) illustrates this situation.
Parameters of the calculation are those of Fig.~\ref{fig_m2_diventra}a
except that $\varepsilon_0=5$~eV. With opening of the inelastic channel
both current and noise grow, while the Fano factor decreases.
Contrary to the situation of resonant tunneling,
asymmetry in the coupling to the leads does not provide qualitative
difference in this case.

\begin{figure}[htpb]
\centering\includegraphics[width=1.2\linewidth]{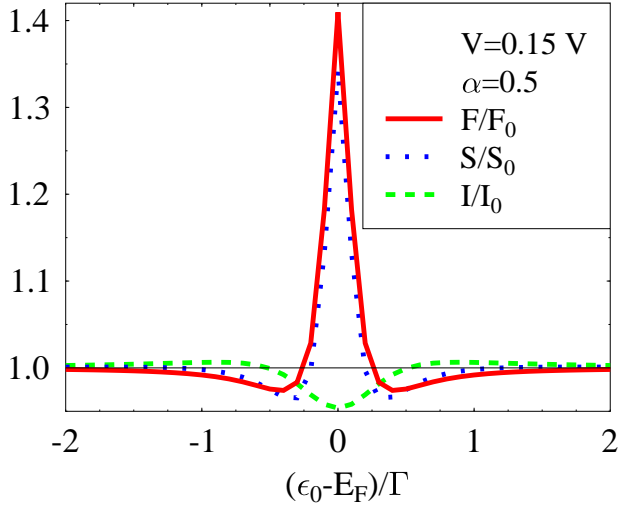}
\caption{\label{fig_m2_e0}(Color online)
Ratios of Fano factors (solid line), zero frequency noises (dotted
line), and currents (dashed line) with and without electron-phonon coupling
vs. position of the molecular orbital relative to the Fermi energy.
See text for parameters.}
\end{figure}

Figure~\ref{fig_m2_e0} shows the effect on the current and noise
properties of transition from the resonant to the off-resonant tunneling 
regime as may be observed by applying a suitable gate potential. 
Parameters of the calculation are those of Fig.~\ref{fig_m2_diventra}a 
with position of the molecular orbital changing relative to the Fermi energy. 
Shown are the ratios of the current (dashed line), noise (dotted line) 
and Fano factor (solid line) to their values
in the absence of electron-phonon interaction.
As noticed above and as implied by Eq.~(\ref{T}) the current goes through
a minimum at resonance. In contrast the noise (and consequently the Fano
factor) maximize at that point. Note that the results are obtained using
$eV=0.15$~eV ${}>\hbar\omega_0$. For $eV<\hbar\omega_0$ phonon effects
are negligible and the resulting lines will approach unity.

\section{\label{strong}Strong electron-phonon coupling}
In this section we sketch the procedure and present results of our numerical 
calculations for the case of intermediate to strong electron-phonon coupling.
Details of the calculation procedure can be found in 
Ref.~\onlinecite{strong_el_ph} and here we mention only several 
important points. A small polaron (canonical or Lang-Firsov) 
transformation\cite{Mahan,Holstein,LangFirsov} applied to 
Hamiltonian~(\ref{H}) leads to
\begin{eqnarray}
 \label{barH}
 \hat{\bar H} &=& \bar\varepsilon_0\hat c^\dagger\hat c +
 \sum_{k\in\{L,R\}} \varepsilon_k \hat c_k^\dagger\hat c_k +
 \sum_{k\in\{L,R\}} \left(V_k\hat c_k^\dagger\hat c\hat X_a+\mbox{h.c.}\right)
 \nonumber \\ &+&
 \omega_0\hat a^\dagger\hat a +
 \sum_\beta\omega_\beta\hat b^\dagger_\beta\hat b_\beta +
 \sum_\beta U_\beta\hat Q_a\hat Q_\beta
\end{eqnarray}
where
\begin{equation}
 \label{bare0}
 \bar\varepsilon_0 = \varepsilon_0 - \Delta \qquad
 \Delta \approx \frac{M_a^2}{\omega_0};
\end{equation}
$\Delta$ is the electron level shift due to coupling to the primary phonon and
\begin{equation}
 \label{Xa}
 \hat X_a = \exp\left[i\lambda_a\hat P_a\right] \qquad
 \lambda_a = \frac{M_a}{\omega_0}
\end{equation}
is the primary phonon shift generator with 
$\hat P_a = -i\left(\hat a - \hat a^\dagger\right)$ 
being the primary phonon momentum.
We are looking for the single electron Green function, approximately
given (on the Keldysh contour) by
\begin{align}
 \label{appGFKeldysh}
 &G(\tau_1,\tau_2) = -\frac{i}{\hbar}
 <T_c \hat c(\tau_1)\hat c^\dagger(\tau_2)>_H
 \nonumber \\
 &\quad= -\frac{i}{\hbar}<T_c \hat c(\tau_1)\hat X_a(\tau_1)\,
          \hat c^\dagger(\tau_2)\hat X_a^\dagger(\tau_2)>_{\bar H}
 \\
 &\quad\approx
 -\frac{i}{\hbar}<T_c \hat c(\tau_1)\hat c^\dagger(\tau_2)>_{\bar H}
 \, <T_c \hat X_a(\tau_1)\hat X_a^\dagger(\tau_2)>_{\bar H}
 \nonumber 
 \\ 
 &\quad\equiv
 G_c(\tau_1,\tau_2)\,<T_c \hat X_a(\tau_1)\hat X_a^\dagger(\tau_2)>_{\bar H}
 \nonumber
\end{align}
In Eq.~(\ref{appGFKeldysh}) $T_c$ is the time ordering operator on the 
Keldysh contour, $\hat{\bar H}$ is the transformed Hamiltonian, 
Eq.~(\ref{barH}), and 
$<\ldots>_{\bar H}$ implies that the indicated time evolutions are to be
carried out with this Hamiltonian. Also
\begin{align}
 \label{XXKeldysh}
 &<T_c \hat X_a(\tau_1)\hat X_a^\dagger(\tau_2)> =
 \\ &\qquad
 \exp\left\{\lambda_a^2\left[i\hbar D_{P_aP_a}(\tau_1,\tau_2)
                -<\hat P_a^2>\right]\right\}
 \nonumber
\end{align}
is an approximate (second order cumulant expansion) expression
for the shift generator correlation function in terms of the primary  
phonon Green function
\begin{equation}
 \label{DKeldysh}
 D_{P_aP_a}(\tau_1,\tau_2) = -\frac{i}{\hbar}
 <T_c \hat P_a(\tau_1)\hat P_a(\tau_2)>
\end{equation}

The (approximate) electron Green function $G$ is obtained from a
self-consistent solution of the approximate (second order in the coupling
to the contacts) equations for the phonon and electron
Green functions
\begin{align}
 \label{Dyson}
 &D_{P_aP_a}(\tau,\tau') = D_{P_aP_a}^{(0)}(\tau,\tau')
 \\
 &+ \int_c d\tau_1 \int_c d\tau_2\, D_{P_aP_a}^{(0)}(\tau,\tau_1)\,
 \Pi_{P_aP_a}(\tau_1,\tau_2)\, D_{P_aP_a}(\tau_2,\tau') 
 \nonumber \\
 \label{GcKeldysh}
 &G_c(\tau,\tau') = G_c^{(0)}(\tau,\tau')
 \\
 &+ \sum_{K=\{L,R\}}\int_c d\tau_1 \int_c d\tau_2\, G_c^{(0)}(\tau,\tau_1)\,
 \Sigma_{c,K}(\tau_1,\tau_2)\, G_c(\tau_2,\tau')
 \nonumber
\end{align}
These equations (for derivation see~\cite{strong_el_ph}) are analogs of the 
usual Dyson equation.
The functions $\Pi_{P_aP_a}$ and $\Sigma_{c,K}$, which are analogs of
usual phonon and electron self-energies, satisfy the equations
\begin{align}
 \label{DSEKeldysh}
 &\Pi_{P_aP_a}(\tau_1,\tau_2) = \sum_\beta |U_\beta|^2
 D_{P_\beta P_\beta}(\tau_1,\tau_2)
 - i\lambda_a^2\sum_{k\in\{L,R\}}|V_k|^2
 \nonumber \\ &\times
 \left[
     \hbar g_k(\tau_2,\tau_1)G_c(\tau_1,\tau_2)
     <T_c \hat X_a(\tau_1)\hat X_a^\dagger(\tau_2)>
 \right. \\  
 &\left.\quad +  (\tau_1\leftrightarrow\tau_2)\right]
  \nonumber \\
 \label{GcSEKeldysh}
 &\Sigma_{c,K}(\tau_1,\tau_2) = \sum_{k\in K} |V_k|^2 g_k(\tau_1,\tau_2)
 <T_c \hat X_a(\tau_2)\hat X_a^\dagger(\tau_1)>
\end{align}
Here $K=L,R$ and $g_k$ is the free electron Green function for state $k$
in the contacts. 

The Langreth projections\cite{Langreth} of 
Eqs.~(\ref{Dyson})-(\ref{GcSEKeldysh})
on the real time axis are solved numerically by using grids in time
and energy spaces, repeatedly switching between these spaces as
described in Ref.~\onlinecite{strong_el_ph}.
After convergence, the Green function and self-energies needed in the 
noise expression (\ref{Sweq0}) are available as numerical functions of $E$.

In the presence of electron-phonon interactions the expression for 
the average steady state current can
still be written in a form similar to (\ref{I0})\cite{current} with
a renormalized transmission coefficient $T(E)$ replacing $T_0(E)$.
$T(E)$ can be written as $T(E)=\Gamma_L\Gamma_R 2\pi\rho_{el}(E)/\Gamma$, 
where $\rho_{el}(E)$ is now the interacting electronic density of 
states.\cite{likeLandauer}
Again, the expression for the noise can no longer be cast in the
simple additive structure of Eq.(\ref{S0}).\cite{additive} 

In what follows we present and discuss some numerical results based on
this procedure. These results will serve to illustrate the sensitivity
of the current and noise calculations to the level of approximation used,
the manifestation of the phonon sideband structure in the noise spectrum and
the conditions for observing a single or split central peak in the noise 
spectrum in strong electron-phonon coupling situations. 
We focus on these issues while limiting the size of our parameter space 
by taking symmetric junctions with 
$\Gamma_L=\Gamma_R$ (i.e. $\alpha=0.5$) and $\eta_l=-\eta_R=0.5$,
$\varepsilon_0=2$~eV above the unbiased Fermi energies, $T=10$~K,
$\omega_0=0.2$~eV, $M_a=0.2$~eV and $\gamma_{ph}=0.01$~eV.

\begin{figure}[t]
\centering\includegraphics[width=\linewidth]{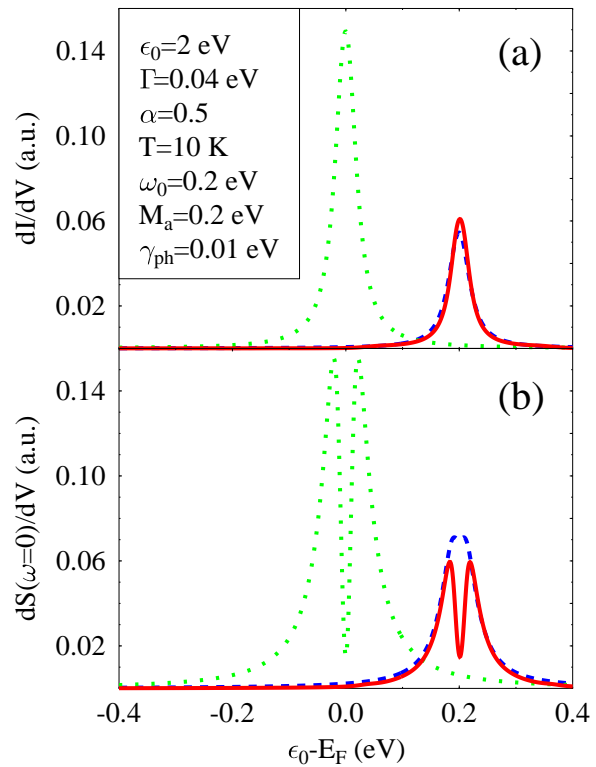}
\caption{\label{fig_noise_sym_eps0}(Color online)
Conductance (a) and differential noise (b) vs. level
position (gate voltage experiment). Shown are results of self-consistent
calculation (solid line) and the zero order approximation (dashed line).
Results in the absence of electron-phonon coupling (dotted line)
are given for comparison.
}
\end{figure}

Figure~\ref{fig_noise_sym_eps0} shows the differential conductance $dI/dV$ and 
noise spectrum $dS/dV$ evaluated at $V=0.01$~V and plotted against 
the resonance energy $\varepsilon_0$ (that is in principle controllable by
an imposed gate potential).  The dotted lines in Fig.~\ref{fig_noise_sym_eps0} 
represent the current and noise obtained in the purely elastic case 
(no electron-phonon interaction).
The solid lines are results of the converged self-consistent procedure
described above while the dashed lines result from the lowest order 
approximation to Eq.~(\ref{appGFKeldysh}) in which the two factors (electronic
and phononic) that enter the Green function $G$ are calculated in the 
absence of electron-phonon interaction.\cite{approx}
Two important features should be noticed.
First, because of its split peak structure, the noise spectrum is far more
sensitive to the electron-phonon coupling than the conduction plot.
Second, for the small source-drain potential used here no phonon sidebands
appear. A similar observation was made before for the conduction 
spectrum.\cite{strong_el_ph,elph_mitra} 
This observation stands in contrast to the results of Zhu and 
Balatsky,\cite{noise_elph_Balatsky} who obtained phonon sideband structure
in the noise spectrum plotted against the gate potential. 
Indeed the procedure of Ref.~\onlinecite{noise_elph_Balatsky} 
would lead to sideband 
structure also in the $dI/dV$ vs. gate potential spectrum and results from 
the semiclassical approximation discussed above.\cite{additive} 
It should be noted that this approximation is
valid only when $\hbar\omega_0\ll k_BT$, whereupon the sideband structure will
be suppressed by thermal broadening associated with the width of the Fermi
function.

\begin{figure}[t]
\centering\includegraphics[width=\linewidth]{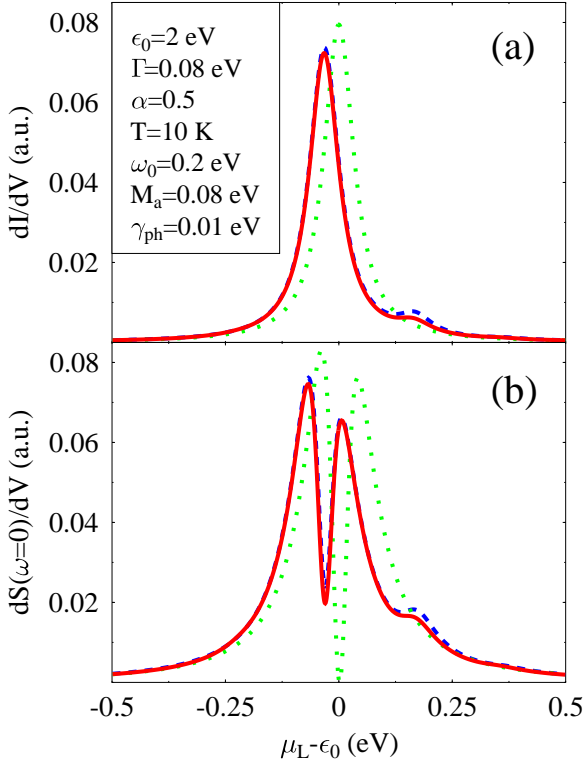}
\caption{\label{fig_noise_balatsky}(Color online)
Conductance (a) and differential noise (b) vs. applied
source-drain voltage for the case of relatively weak electron-phonon coupling.
The different lines represent results for the full and zero order
calculation and for the case with no electron-phonon coupling (line notation
as in Fig.~\ref{fig_noise_sym_eps0}).
}
\end{figure}

In contrast to gate-voltage plots made at small bias potentials,
the phonon sideband structure is observed when the
differential conductance is plotted against the source-drain voltage 
$V$.\cite{strong_el_ph,elph_mitra}
A similar behavior is observed for the noise spectrum.
Figure~\ref{fig_noise_balatsky}
presents results of the corresponding calculation for the case of relatively
weak electron-phonon coupling. Parameters of the calculation are 
the same as in Fig.~\ref{fig_noise_sym_eps0} 
except $\Gamma_L=\Gamma_R=0.08$~eV and $M_a=0.08$~eV. 
In this weak coupling case only one sideband appears, indicating a 
single phonon creation by the tunneling electron. Also in this limit
the low order calculation appears to be sufficient --- self-consistent
corrections are seen to be small.

\begin{figure}[t]
\centering\includegraphics[width=\linewidth]{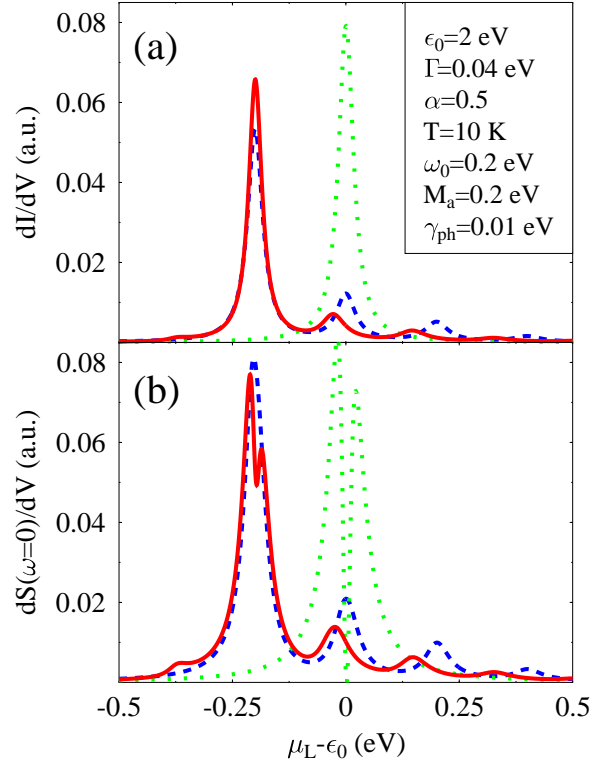}
\caption{\label{fig_noise_sym}(Color online)
Conductance (a) and differential noise (b) vs. applied
source-drain voltage for the case of relatively strong electron-phonon coupling.Line notation is as in Figs.~\ref{fig_noise_sym_eps0} and
\ref{fig_noise_balatsky}.
}
\end{figure}

A stronger coupling case is presented in Figure~\ref{fig_noise_sym}, where the 
parameters of calculation are the same as in Fig~\ref{fig_noise_sym_eps0}.
Several phonon sidebands indicating phonon emission are observed 
on the right of the central (elastic) peak. The satellite feature that
appears to the left of the central peak is obtained
only in the self-consistent calculation (solid line).
As was discussed in~\onlinecite{strong_el_ph} this feature indicates
phonon absorption by the tunneling electron and results from heating
of the junction by electron flux. We see again that the shape of the 
differential noise curve, Fig.~\ref{fig_noise_sym}b, is more sensitive to 
interaction renormalization (self-consistency of calculation) than that of 
the conductance, Fig~\ref{fig_noise_sym}a, especially in its central peak 
structure.

A significant difference between the split peak structure of the elastic
central peak in Figures~\ref{fig_noise_balatsky} and \ref{fig_noise_sym} 
(solid lines) and the elastic result (Fig.~\ref{fig_noise0_a} and dotted lines 
of Figs.~\ref{fig_noise_balatsky} and \ref{fig_noise_sym}) 
when considered in the symmetric molecule-leads coupling
case, $\alpha=0.5$, is the fact that the differential noise spectrum 
vanishes between the two peaks in the absence of electron-phonon coupling
but remains finite when this coupling is present. To examine the
significance of this difference we focus on the $T=0$ limit and 
the zero order approximation.
The lesser and greater Green functions are given by
\begin{align}
 \label{Glt_T0}
 &G^{<}(E) = i e^{-\lambda_a^2} \sum_{n=0}^{\infty}\frac{\lambda_a^{2n}}{n!}
 \\ &\times
 \frac{\Gamma_L\theta(\mu_L-E-n\omega_0)+\Gamma_R\theta(\mu_R-E-n\omega_0)}
      {(E-\varepsilon_0+n\omega_0)^2+(\Gamma/2)^2} 
 \nonumber \\
 \label{Ggt_T0}
 &G^{>}(E) = -i e^{-\lambda_a^2} \sum_{n=0}^{\infty}\frac{\lambda_a^{2n}}{n!}
 \\ &\times
 \frac{\Gamma_L\theta(E-n\omega_0-\mu_L)+\Gamma_R\theta(E-n\omega_0-\mu_R)}
      {(E-\varepsilon_0-n\omega_0)^2+(\Gamma/2)^2}
 \nonumber
\end{align}

The retarded Green function can be estimated from the Lehmann 
representation\cite{Mahan}
\begin{equation}
 \label{Lehmann}
 G^r(E) = i \int_{-\infty}^{+\infty}\frac{dE'}{2\pi}\,
 \frac{G^{>}(E')-G^{<}(E')}{E-E'+i\delta}
\end{equation}
For the case of relatively weak coupling to the leads $\Gamma\ll\omega_0$ 
(which is the case where phonon sidebands may be observed), the part of $G^r$ 
responsible for the central peak region can be approximated by using in 
(\ref{Lehmann}) only the $n=0$ terms of (\ref{Glt_T0}) and (\ref{Ggt_T0}).
This leads to 
$G^r_{n=0}(E)=e^{-\lambda_a^2}[E-\varepsilon_0+i\Gamma/2]^{-1}$.
Using this expression for $G^r$ together with $G^{<}$ and $G^{>}$
approximated again by their $n=0$ terms in Eq.~(\ref{Sweq0}) leads to
\begin{widetext}
\begin{align}
 \label{noise_central_region}
 \frac{dS_{n=0}(\omega=0)}{dV} &= \frac{4e^3}{2\pi\hbar}
 \left[\delta\times T_0(\mu_L)e^{-\lambda_a^2}
      \left(\eta_L^2+\eta_R^2-2\eta_L\eta_Re^{-\lambda_a^2}
      -T_0(\mu_L)e^{-\lambda_a^2}\right)\right.
 \\ &\quad
 \left. + (1-\delta)\times T_0(\mu_R)e^{-\lambda_a^2}
      \left(\eta_L^2+\eta_R^2-2\eta_L\eta_Re^{-\lambda_a^2}
      -T_0(\mu_R)e^{-\lambda_a^2}\right) \right]
 \nonumber
\end{align}
\end{widetext}
where we have also used Eqs.~(\ref{SEKlt})-(\ref{GammaK}) and where 
$T_0$ is given by Eq.~(\ref{T0}).

Eq.~(\ref{noise_central_region}) is the analog of Eq.~(\ref{dSs_dV})
(the thermal noise contribution vanishes in this $T=0$ limit),
and indeed leads to (\ref{dSs_dV}) for $\lambda_a=0$. For finite
electron-phonon coupling the result is seen to depend on the capacitance
ratios $\eta_L$ and $\eta_R$. The requirement that $T_0(E)[1-T_0(E)]$
has three extrema for a double-peak structure to be seen in the 
$dS/dV$ vs. $V$ plot can now be applied for the function 
$T_0(E)e^{-\lambda_a^2}\left(\eta_L^2+\eta_R^2-2\eta_L\eta_Re^{-\lambda_a^2}
-T_0(E)e^{-\lambda_a^2}\right)$. This leads to an analog of 
Eq.~(\ref{cond_dp_0})
\begin{align}
 \label{cond_dp}
 &\alpha^2-\alpha+\frac{p}{8} < 0 \\
 \label{p}
 & p \equiv e^{\lambda_a^2} + 2\eta_L\eta_R\left(e^{\lambda_a^2}-1\right)
\end{align}
which reduces to (\ref{cond_dp_0}) when $\lambda_a\to 0$.
The peak positions are given by 
$\varepsilon_0\mp[\Gamma/2]\sqrt{8\alpha(1-\alpha)/p-1}$.
One can show also that
\begin{align}
 \label{noise_peak}
 &\left.\frac{dS}{dV}\right|_{peak} = \frac{4e^3}{2\pi\hbar}\delta\,
 e^{-2\lambda_a^2}\,\frac{p^2}{4}
 \\
 \label{noise_e0}
 &\left.\frac{dS}{dV}\right|_{\varepsilon_0} = 
 \frac{4e^3}{2\pi\hbar}\delta\,e^{-2\lambda_a^2}\,
 16\alpha(1-\alpha)[p/4-\alpha(1-\alpha)]
\end{align}
In the absence of electron-phonon coupling $p=1$ and the expressions
above take the forms shown in section~\ref{elastic} for elastic tunneling
case, where for $\alpha=0.5$ $\left.\frac{dS}{dV}\right|_{\varepsilon_0}=0$.
This vanishing does not occur even for $\alpha=0.5$ if $p\neq 1$.
Consequently, the peak positions and noise values spectrum near and between 
the peaks contains information about the asymmetry of the coupling to the 
leads, capacitance parameters and strength of electron-phonon interaction.

\begin{figure}[t]
\centering\includegraphics[width=\linewidth]{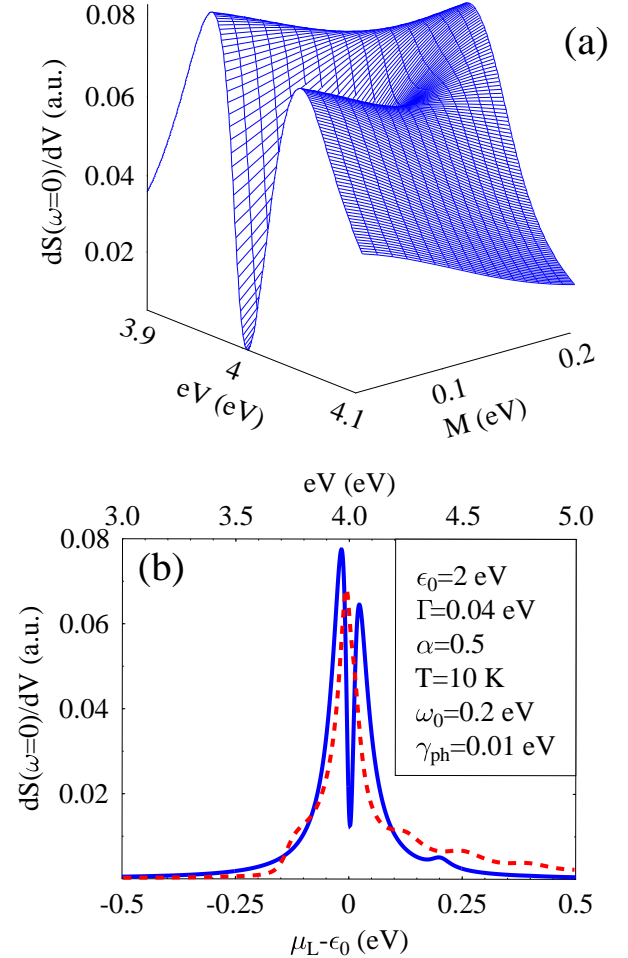}
\caption{\label{fig_noise_m}(Color online) 
Differential noise vs. applied
source-drain voltage for $\alpha=0.5$.
(a) Surface plot of $dS(\omega=0)/dV$ as function of source-drain voltage and
 electron-phonon coupling strength. Zero-order result.
(b) $dS(\omega=0)/dV$ for weak ($M=0.04$~eV, solid line) 
and strong ($M=0.3$~eV, dashed line) electron-phonon coupling.
Self-consistent calculation.
}
\end{figure}

For very strong electron-phonon interaction, 
$\lambda_a\to\infty$, condition (\ref{cond_dp}) can not be satisfied 
for any choice of the junction parameters. In this case only one peak 
in $dS/dV$ in the central region will be observed. 
Figure~\ref{fig_noise_m} illustrates this effect. 
The parameters used in calculation are the same as in Fig.~\ref{fig_noise_sym}. 
The calculation is done for symmetric coupling, $\alpha=0.5$ and $\eta=0.5$.  
This behavior is seen already in the zero order approximation as is shown 
in Fig.~\ref{fig_noise_m}.
Fig.~\ref{fig_noise_m}a shows evolution of the two peak structure into the one 
central peak with increase of electron-phonon coupling, 
calculated in this level of approximation.
Fig.~\ref{fig_noise_m}b shows result of self-consistent calculation.
A two peak structure is observed for weak 
($M_a=0.04$~eV, solid line) and a single peak for strong ($M_a=0.3$~eV, 
dashed line) electron-phonon coupling.

This spectral change from two to one peak structure with increasing
electron-phonon coupling results from transient localization and
increasing dephasing of the electron on the bridge and can be used for 
estimating the junction parameters.
For example, in an STM experiment, increasing tip-substrate
distance will lead (for not too strong electron-phonon coupling)
to a change of the differential noise/voltage profile from a two to a one-peak
structure as was seen in Fig.~\ref{fig_noise0_a} 
(the general shape of Fig.~\ref{fig_noise0_a}
will persist also for a non-zero $M_a$). Eq.~(\ref{cond_dp}) written as an
equality is a relationship between $\alpha$, $\eta$, $\omega_0$ and $M_a$
at this transition point. 
Information on $\omega_0$ is easily obtained from the inelastic tunneling
structure of either the average current or the noise.
To get information on $\alpha$ one can use Eqs.~(\ref{IK}), (\ref{Glt_T0}) and 
(\ref{Ggt_T0}) to get for $T\to 0$
\begin{align}
 \label{dI_dV}
 &\frac{dI}{dV} = \frac{2e^2}{2\pi\hbar}\frac{\Gamma_L\Gamma_R}{\Gamma}
 e^{-\lambda_a^2}\sum_{n=0}^\infty \frac{\lambda_a^{2n}}{n!}
 \nonumber \\ &\times
 \left\{\theta(eV-n\omega_0)\left[
 \frac{\delta}{\Gamma_L}T_{+n}(\mu_L)+\frac{1-\delta}{\Gamma_R}T_{-n}(\mu_R)
 \right]\right.
 \\
 & \left.+\theta(-eV-n\omega_0)\left[
 \frac{\delta}{\Gamma_L}T_{-n}(\mu_L)+\frac{1-\delta}{\Gamma_R}T_{+n}(\mu_R)
 \right]\right\}
 \nonumber
\end{align}
where 
\begin{equation}
 \label{Tn}
 T_{\pm n}(E) = \frac{\Gamma_L\Gamma_R}
                {(E-\varepsilon_0\mp n\omega_0)^2+(\Gamma/2)^2}
\end{equation}
It is easy to see from (\ref{dI_dV}) and (\ref{Tn}) that 
\begin{equation}
 \frac{\left.\frac{dI}{dV}\right|_{\mu_R=\varepsilon_0} \times
       \left.\frac{dI}{dV}\right|_{\mu_L=\varepsilon_0+n\omega_0}}
      {\left.\frac{dI}{dV}\right|_{\mu_L=\varepsilon_0} \times
       \left.\frac{dI}{dV}\right|_{\mu_R=\varepsilon_0+n\omega_0}}
       = \frac{1-\alpha}{\alpha}
\end{equation}
In addition Eq.~(\ref{params}) provides a relationship between the parameters
$\alpha$, $\delta$ and $\eta$. An independent determination of the voltage
division factor $\delta$ will therefore determine also the other parameters
(within our simple junction model). We note that information on the voltage
distribution in a junction can be obtained directly\cite{McEuen}
or indirectly from $I/V$ plots at positive and negative bias.

\section{\label{conclude}Conclusion}
We have studied inelastic effects on the conduction and noise properties of 
molecular junctions within a simple model that comprises a one level bridge
(arbitrarily chosen above as the molecular LUMO) between two metal contacts.
The level is coupled to a vibrational degree of freedom that represents
a molecular vibration, which in turn is coupled to a thermal phonon bath
that represents the environment.

The case of weak electron-phonon coupling is investigated using 
standard diagrammatic technique on the Keldysh contour as 
described in~\onlinecite{IETS_SCBA}. We find that the
predictions of Ref.~\onlinecite{t_stmol_DiVentra}, 
increase in zero frequency noise
and Fano factor with opening of the inelastic channel, hold only in a
particular range of parameters --- resonant tunneling with symmetric coupling
to the leads. Different modes of behavior are found upon changes that can be 
affected by moving an STM tip or changing the gate potential.
The crossover point between the different modes of behavior 
provides information on the strength and asymmetry of the molecule-leads 
coupling and the position of the molecular level relative to the Fermi energy.

For the moderate to strong electron-phonon coupling case we have
implemented a recently developed self-consistent scheme on the Keldysh
contour.\cite{strong_el_ph} We have studied the zero-frequency 
differential noise spectrum for resonant inelastic tunneling and have
investigated the shape of this spectrum and its dependence on the
parameters of the junction (asymmetry in coupling to the leads and
electron-phonon interaction strength). 
We find that 
\begin{itemize}
\item As in the $I/V$ spectrum, the differential noise spectrum in
resonance inelastic tunneling shows a typical structure characterized
by a central feature and phonon sidebands.
\item Contrary to Ref.~\cite{noise_elph_Balatsky} we find that 
displaying the differential noise $dS(\omega=0)/dV$ against the gate potential
does not allow observation of phonon sidebands.
Rather the sideband structure can be observed in the usual source-drain
measurements.
\item The central feature exhibits a crossover from double to one central 
peak structure, both for increasing asymmetry in molecule-lead coupling 
and for increasing vibronic coupling strength.
This change of spectral shape may be used to gain information on the 
parameters that determine these properties.
In particular, asymmetry can be controlled by the tip-molecule distance in
STM experiments.
\end{itemize}
It should be emphasized that as in other studies of current noise in
nanojunctions we have focused on the electronic current in static
junction structures, here augmented by electron-phonon interactions.
Other sources of noise, e.g. structural fluctuations that affect tunneling
matrix elements and molecule-lead coupling may exist, 
and future studies should assess their potential contributions.

\begin{acknowledgments}
MR thanks Mark Reed for a fruitful discussion. 
We thank Yu.~Galperin for clarifying for us the origin of Eq.~(\ref{hatI}).
We thank the DARPA MoleApps program and the NASA URETI program for support.
AN thanks the Israel Science Foundation and the U.S.-Israel Binational
Science Foundation for support.
\end{acknowledgments}


\begin{thebibliography}{99}
\bibitem{MEB}{\em Molecular Electronics II}, Eds. A.~Aviram, M.~Ratner,
   V.~Mujica, Ann.~N.Y.~Acad.~Sci., vol. 960 (2002);\\
   {\em Molecular Electronics III}, Eds. J.~R.~Reimers, C.~A.~Picconatto,
   J.~C.~Ellenbogen, R.~Shashidhar, Ann.~N.Y.~Acad.~Sci., vol. 1006 (2003).
\bibitem{AN}A.~Nitzan, Ann.~Rev.~Phys.~Chem. \textbf{53}, 681 (2001).
\bibitem{Reed}{\em Molecular Nanoelectronics}, Eds. M.~A.~Reed and T.~Lee,
   American Scientific Publishers (2003);\\
   W.~Y.~Wang and M.~A.~Reed, Rep.~Progr.~Phys. \textbf{68}, 523 (2005).
\bibitem{Cuniberti}{\em Introducing Molecular Electronics},
   Eds. G.~Cuniberti, G.~Fagas, and K.~Richter, 
   Springer, Berlin and Heidelberg (2005).
\bibitem{Bowler}D.~R.~Bowler,
   J.~Phys.: Condens.~Matter \textbf{16}, R721 (2004).
\bibitem{lang} A.~Yazdani, D.~M.~Eigler, and N.~D.~Lang,
   Science \textbf{272}, 1921 (1996).
\bibitem{kubiak}R.~P.~Andres, T.~Bein, M.~Dorogi, S.~Feng, J.~I.~Henderson,
   C.~P.~Kubiak, W.~Mahoney, R.~G.~Osifchin, R.~Reifenberger,
   Science \textbf{272}, 1323 (1996);
   S.~Datta, W.~Tian, S.~Hong, R.~Reifenberger, J.~I.~Henderson,
   and C.~P.~Kubiak, Phys.~Rev.~Lett. \textbf{79}, 2530 (1997);
   W.~Tian, S.~Datta, S.~Hong, R.~Reifenberger, J.~I.~Henderson,
   and C.~P.~Kubiak, J.~Chem.~Phys. \textbf{109}, 2874 (1998).
\bibitem{reed_tour}M.~A.~Reed, C.~Zhou, C.~J.~Muller, T.~P.~Burgin,
   and J.~M.~Tour, Science \textbf{278}, 252 (1997);
   C.~Zhou, M.~R.~Deshpande, M.~A.~Reed, L.~Jones~II, and J.~M.~Tour,
   Appl.~Phys.~Lett. \textbf{71}, 611 (1997); 
   ibid. \textbf{71}, 2857 (1997);\\
   J.~Chen, M.~A.~Reed, A.~M.~Rawlett, J.~M.~Tour,
   Science \textbf{286}, 1550 (1999).
\bibitem{heath}C.~P.~Collier, E.~W.~Wong, M.~Belohradsky, F.~M.~Raymo,
   J.~F.~Stoddart, P.~J.~Kuekes, R.~S.~Williams, and J.~R.~Heath,
   Science \textbf{285}, 391 (1999).
\bibitem{joachim} C.~Kergueris, J.-P.~Bourgoin, S.~Palacin, D.~Esteve,
   C.~Urbina, M.~Magoga, and C.~Joachim,
   Phys.~Rev.~B \textbf{59}, 12505 (1999).
\bibitem{weber}J.~Reichert, R.~Ochs, D.~Beckman, H.~B.~Weber, M.~Mayor,
   and H.~v.~L\"ohneysen, Phys.~Rev.~Lett. \textbf{88}, 176804 (2002).
\bibitem{park}W.~Liang, M.~P.~Shores, M.~Bockrath, J.~R.~Long, and H.~Park,
   Nature \textbf{417}, 725 (2002).
\bibitem{bjornholm}S.~Kubatkin, A.~Danilov, M.~Hjort, J.~Cornil, J.-L.~Bredas,
   N.~Stuhr-Hansen, P.~Hedegard, and T.~Bj\o rnholm,
   Nature \textbf{425}, 698 (2003).
\bibitem{mceuen}H.~Park, J.~Park, A.~K.~L.~Lim, E.~H.~Anderson, 
   A.~P.~Alivisatos, and P.~L.~McEuen, Nature \textbf{407}, 57 (2000);\\
   J.~Park, A.~N.~Pasupathy, J.~I.~Goldsmith, C.~Chang, Y.~Yaish, J.~R.~Petta,
   M.~Rinkoski, J.~P.~Sethna, H.~D.~Abru\~na, P.~L.~McEuen, and D.~C.~Ralph,
   Nature \textbf{417}, 722 (2002);\\
   J.~Park, A.~N.~Pasupathy, J.~I.~Goldsmith, A.~V.~Soldatov, C.~Chang,
   Y.~Yaish, J.~P.~Sethna, H.~D.~Abru\~na, D.~C.~Ralph, and P.~L.~McEuen,
   Thin Solid Films \textbf{438-439}, 457 (2003);\\
   A.~N.~Pasupathy, R.~C.~Bialczak, J.~Martinek, J.~E.~Grose, L.~A.~K.~Donev,
   P.~L.~McEuen, and D.~C.~Ralph, Science \textbf{306}, 86 (2004).
\bibitem{natelson}L.~H.~Yu, Z.~K.~Keane, J.~W.~Ciszek, L.~Cheng, M.~P.~Stewart,
   J.~M.~Tour, and D.~Natelson, Phys.~Rev.~Lett. \textbf{93}, 266802 (2004);\\
   L.~H.~Yu and D.~Natelson, Nano~Lett. \textbf{4}, 79 (2004);\\
   L.~H.~Yu, Z.~K.~Keane, J.~W.~Ciszek, L.~Cheng, J.~M.~Tour, T.~Baruah,
   M.~R.~Pederson, and D.~Natelson, cond-mat/0505683 (2005).
\bibitem{ho}H.~J.~Lee and W.~Ho, Science \textbf{286}, 1719 (1999);
   Phys.~Rev.~B \textbf{61}, R16347 (2000);\\
   J.~Gaudioso, L.~J.~Lauhon, and W.~Ho, Phys.~Rev.~Lett. \textbf{85}, 1918 (2000);\\
   J.~R.~Hahn, H.~J.~Lee, and W.~Ho, Phys.~Rev.~Lett. \textbf{85}, 1914 (2000);\\
   L.~J.~Lauhon and W.~Ho, Phys.~Rev.~Lett. \textbf{85}, 4566 (2000);
   Phys.~Rev.~B \textbf{60}, R8525 (1999);\\
   N.~Lorente, M.~Persson, L.~J.~Lauhon, and W.~Ho, Phys.~Rev.~Lett. \textbf{86}, 2593 (2001);\\
   J.~R.~Hahn and W.~Ho, Phys.~Rev.~Lett. \textbf{87}, 196102 (2001).
\bibitem{zhitenev}N.~B.Zhitenev, H.~Meng, and Z.~Bao,
   Phys.~Rev.~Lett. \textbf{88}, 226801 (2002).
\bibitem{ruitenbeek}R.~H.~M.~Smit, Y.~Noat, C.~Untiedt, N.~D.~Lang, 
   M.~C.~van~Hemert, and J.~M.~van~Ruitenbeek,
   Nature \textbf{419}, 906 (2002);\\
   D.~Djukic, K.~S.~Thygesen, C.~Untiedt, R.~H.~M.~Smit, K.~W.~Jacobsen,
   and J.~M.~van~Ruitenbeek,
   Phys.~Rev.~B \textbf{71}, 161402(R) (2005).
\bibitem{noiserev_Buttiker} Ya.~M.~Blanter and M.~Buttiker,
   Physics Reports \textbf{336}, 1 (2000).
\bibitem{e_DBRTS_Li} Y.~P.~Li, A.~Zaslavsky, D.~C.~Tsui, M.~Santos, 
   and M.~Shayegan, Phys.~Rev.~B \textbf{41}, 8388 (1990).
\bibitem{e_DBRTS_Safonov} S.~S.~Safonov, A.~K.~Savchenko, D.~A.~Bagrets,
   O.~N.~Jouravlev, Y.~V.~Nazarov, E.~H.~Linfield, and D.~A.~Ritchie,
   Phys.~Rev.~Lett. \textbf{91}, 136801 (2003).
\bibitem{e_DBRTS_Nauen} A.~Nauen, F.~Hohls, N.~Maire, K.~Pierz, and R.~J.~Haug,
   Phys.~Rev.~B \textbf{70}, 033305 (2004).
\bibitem{e_DBRTS_Jung} S.~W.~Jung, T.~Fujisawa, Y.~Hirayama, and Y.~H.~Jeong,
   Appl.~Phys.~Lett. \textbf{85}, 768 (2004).
\bibitem{e_QPC_Li} Y.~P.~Li, D.~C.~Tsui, J.~J.~Heremans, J.~A.~Simmons, 
   and G.~W.~Weimann, Appl.~Phys.~Lett. \textbf{57}, 774 (1990).
\bibitem{e_QPC_Washburn} S.~Washburn, R.~J.~Haug, K.~Y.~Lee, and J.~M.~Hong,
   Phys.~Rev.~B \textbf{44}, 3875 (1991).
\bibitem{e_QPC_Liefrink} F.~Liefrink, J.~I.~Dijkhuis, and H.~van~Houten,
   Semicond.~Sci.~Technol. \textbf{9}, 2178 (1994).
\bibitem{e_QPC_Reznikov} M.~Reznikov, M.~Heiblum, H.~Strikman, and D.~Mahalu,
   Phys.~Rev.~Lett. \textbf{75}, 3340 (1995).
\bibitem{e_JJ_Delahaye} J.~Delahaye, R.~Lindell, M.~S.~Silanpaa, M.~A.~Paalanen,
   E.~B.~Sonin, and P.~J.~Hakonen, cond-mat/0209076 (2002).
\bibitem{e_JJ_Lindell} R.~K.~Lindell, J.~Delahaye, M.~A.~Sillanpaa, 
   T.~T.~Heikkila, E.~B.~Sonin, and P.~J.~Hakonen,
   Phys.~Rev.~Lett. \textbf{93}, 197002 (2004).
\bibitem{e_3rdmoment_Reulet} B.~Reulet, J.~Senzier, and D.~E.~Prober,
   Phys.~Rev.~Lett. \textbf{91}, 196601 (2003).
\bibitem{t_DBRTS_Chen} L.~Y.~Chen and C.~S.~Ting,
   Phys.~Rev.~B \textbf{43}, 4534 (1991).
\bibitem{t_DBRTS_Levy} A.~Levy~Yeyati, F.~Flores, and E.~V.~Anda,
   Phys.~Rev.~B \textbf{47}, 10543 (1993).
\bibitem{t_DBRTS_Hung} K.-M.~Hung and G.~Y.~Wu,
   Phys.~Rev.~B \textbf{48}, 14687 (1993).
\bibitem{t_DBRTS_Thielmann} A.~Thielmann, M.~H.~Hettler, J.~K\"onig, and
   G.~Sch\"on, Phys.~Rev.~B \textbf{68}, 115105 (2003).
\bibitem{t_QPC_Averin} D.~Averin and H.~T.~Imam,
   Phys.~Rev.~Lett. \textbf{76}, 3814 (1996).
\bibitem{t_QPC_Green} F.~Green, J.~S.~Thakur, and M.~P.~Das,
   Phys.~Rev.~Lett. \textbf{92}, 156804 (2004).
\bibitem{t_JJ_Sonin2} E.~B.~Sonin,
   Phys.~Rev.~B \textbf{70}, 140506 (2004); cond-mat/0505424 (2005).
\bibitem{t_chain_Aghassi} J.~Aghassi, A.~Thielmann, M.~H.~Hettler,
   and G.~Sch\"on, cond-mat/0505345 (2005).
\bibitem{t_chain_Kinderman} M.~Kindermann and P.~W.~Brouwer,
   cond-mat/0506455 (2005).
\bibitem{t_disordered_Gutman} D.~B.~Gutman and Y.~Gefen,
   Phys.~Rev.~B \textbf{64}, 205317 (2001).
\bibitem{t_mol_Dallakyan} S.~Dallakyan and S.~Mazumdar,
   Appl.~Phys.~Lett. \textbf{82}, 2488 (2003).
\bibitem{t_mol_Walczak} K.~Walczak,
   Phys.~Stat.~Sol~(b) \textbf{241}, 2555 (2004). 
\bibitem{t_mol_DiVentra2} Y.-C.~Chen and M.~Di~Ventra,
   Phys.~Rev.~B \textbf{67}, 153304 (2003);
   J.~Lagerqvist, Y.-C.~Chen, and M.~Di~Ventra,
   Nanotechnology \textbf{15}, S459 (2004).
\bibitem{t_NEMS_Nishiguchi} N.~Nishiguchi,
   Phys.~Rev.~Lett. \textbf{89}, 066802 (2002).
\bibitem{t_NEMS_Smirnov} A.~Yu.~Smirnov, L.~G.~Mourokh, and N.~J.~M.~Horing,
   Phys.~Rev.~B \textbf{67}, 115312 (2003).
\bibitem{t_NEMS_Clerk} A.~A.~Clerk and S.~M.~Girvin,
   Phys.~Rev.~B \textbf{70}, 121303 (2004).
\bibitem{t_NEMS_Novotny} T.~Novotn\'y, A.~Donarini, C.~Flindt, and A.-P.~Jauho,
   Phys.~Rev.~Lett. \textbf{92}, 248302 (2004).
\bibitem{t_NEMS_Flindt} C.~Flindt, T.~Novotny, and A.-P.~Jauho,
   Phys.~Rev.~B \textbf{70}, 205334 (2004).
\bibitem{t_NEMS_Wabnig} J.~Wabnig, D.~V.~Khomitsky, J.~Rammer, and 
   A.~L.~Shelankov, cond-mat/0506802 (2005).
\bibitem{t_AC_Camalet} S.~Camalet, S.~Kohler, and P.~H\"anggi,
   Phys.~Rev.~B \textbf{70}, 155326 (2004).
\bibitem{t_AC_Guyon} R.~Guyon, T.~Jonckheere, V.~Mujica, A.~Cr\'epieux, 
   and T.~Martin, J.~Chem.~Phys. \textbf{122}, 144703 (2005).
\bibitem{t_stmol_Shimizu} A.~Shimizu and M.~Ueda,
   Phys.~Rev.~Lett. \textbf{69}, 1403 (1992).
\bibitem{t_stmol_Bo} O.~Lund Bo and Yu.~Galperin,
   Phys.~Rev.~B \textbf{55}, 1696 (1997).
\bibitem{t_stmol_Dong} B.~Dong, H.~L.~Cui, X.~L.~Lei, and N.~J.~M.~Horing,
   Phys.~Rev.~B \textbf{71}, 045331 (2005).
\bibitem{t_stmol_DiVentra} Y.-C.~Chen and M.~Di~Ventra,
   Phys.~Rev.~Lett. \textbf{95}, 166802 (2005).
\bibitem{noise_elph_Balatsky} J.-X.~Zhu and A.~V.~Balatsky,
   Phys.~Rev.~B \textbf{67}, 165326 (2003).
\bibitem{IETS_SCBA}M.~Galperin, M.~A.~Ratner, and A.~Nitzan,
   J.~Chem.~Phys. \textbf{121}, 11965 (2004).
\bibitem{strong_el_ph} M.~Galperin, A.~Nitzan, and M.~A.~Ratner,
   Phys.~Rev.~B \textbf{73}, 082604 (2006).
\bibitem{weakM}For the resonance tunneling problem under discussion
   the weak and strong electron-phonon coupling limit may be characterized
   by comparing the electron-phonon coupling $M$ to the magnitude of the
   complex resonance energy $\sqrt{\Delta E^2+\Gamma^2/4}$.
\bibitem{Lundin}U.~Lundin and R.~H.~McKenzie,
   Phys.~Rev.~B \textbf{66}, 075303 (2002).
\bibitem{Bickers}N.~E.~Bickers, 
   Rev.~Mod.~Phys. \textbf{59}, 845 (1987).
\bibitem{current}Y.~Meir and N.~S.~Wingreen,
   Phys.~Rev.~Lett. \textbf{68}, 2512 (1992);\\
   A.~Jauho, N.~S.~Wingreen, and Y.~Meir,
   Phys.~Rev.~B \textbf{50}, 5528 (1994).
\bibitem{HaugJauho}H.~Haug and A.-P.~Jauho,
   {\em Quantum Kinetics in Transport and Optics of Semiconductors.}
   (Springer, Berlin, 1996).
\bibitem{Galperin}U.~Hanke, Yu.~Galperin, K.~A.~Chao, M.~Gisself\"alt,
   M.~Jonson, and R.~I.~Shekhter,
   Phys.~Rev.~B \textbf{51}, 9084 (1995).
\bibitem{noise_Galperin} O.~Lund~Bo and Yu.~Galperin, 
   J.~Phys.: Condens.~Matter \textbf{8}, 3033 (1996).
\bibitem{noise_expression}Note that Bo and Galperin\cite{noise_Galperin}
   used a different Green functions notaions. 
\bibitem{Ho}N.~A.~Pradhan, N.~Liu, and W.~Ho,
   J.~Phys.~Chem.~B \textbf{109}, 8513 (2005).
\bibitem{Repp}J.~Repp, G.~Meyer, S.~M.~Stojkovic, A.~Gourdon, and C.~Joachim,
   Phys.~Rev.~Lett. \textbf{94}, 026803 (2005).
\bibitem{Tinkham}A.~E.~Hanna and M.~Tinkham,
   Phys.~Rev.~B \textbf{44}, 5919 (1991).
\bibitem{IngoldNazarov}G.-L.~Ingold and Yu.~V.~Nazarov in
   {\em Single Charge Tunneling}, edited by H.~Grabert and M.~H.~Devoret
   (Plenum Press, New~York and London, 1992) pp.~21-108.
\bibitem{Silbey}R.~Chance, A.~Prock, and R.~Silbey,
   Adv.~Chem.~Phys. \textbf{31}, 1 (1978).
\bibitem{McEuen}A.~Bachtold, M.~S.~Fuhrer, S.~Plyasunov, M.~Forero,
   E.~H.~Anderson, A.~Zettl, and P.~L.~McEuen,
   Phys.~Rev.~Lett. \textbf{84}, 6082 (2000).
\bibitem{Mahan}G.~D.~Mahan, {\em Many-Particle Physics.} (Third edition,
   Kluwer Academic/Plenum Publishers, New~York, 2000).
\bibitem{Holstein}T.~Holstein, Ann.~Phys. \textbf{8}, 343 (1959).
\bibitem{LangFirsov} I.~G.~Lang and Yu.~A.~Firsov,
   Sov.~Phys.~JETP \textbf{16}, 1301 (1963).
\bibitem{Langreth} D.~C.~Langreth, {\em Linear and Nonlinear Response Theory
   with Applications\/}, p.~3--32 in: {\em Linear and Nonlinear Electron
   Transport in Solids}, Eds. J.~T.~Devreese and D.~E.~Doren
   (Plenum Press, New~York and London, 1976).
\bibitem{muRHOMO}The alternative scenario in which $\mu_R$ crosses the HOMO
for increasingly positive bias would lead to a similar picture for the same 
electron-phonon coupling.
\bibitem{heightasym}Note that asymmetry in these heights may reflect deviations
   from the wide band approximation.
\bibitem{occ}$n(E)$ and $N(\omega)$ are related to the electron and phonon 
   Green functions by $G^{<}(E)=2\pi i\rho_{el}(E)n(E)$ and 
   $D^{<}(\omega)=-2\pi i\rho_{ph}(\omega)\left\{\begin{array}{ll}N(\omega)&\omega>0\\ N(|\omega|)+1&\omega<0\end{array}\right.$.
\bibitem{likeLandauer}This statement holds in the wide band limit or
   at least when $\Gamma_L(E)/\Gamma_R(E)$ does not depend on 
   $E$.\cite{current,HaugJauho} Note that in a more general formulation
   (e.g. for a bridge with more than one level) $2\pi\rho_{el}(E)$
   should be substituted by $A(E)$, where $A(E)$ is a spectral function. 
\bibitem{additive}Previous statements about the additive structure of the
   noise spectrum in the presence of electron-phonon 
   interaction\cite{noise_elph_Balatsky} are based on a treatment that 
   makes the classical-like assumption about the nuclear motion, i.e.
   disregarding the difference between $<\hat X_a(t)\hat X_a^\dagger(0)>$
   and $<\hat X_a^\dagger(0)\hat X_a(t)>$ which is valid only when
   $\hbar\omega_0\ll k_BT$. The ansatz proposed by Ng\cite{ac_current_Ng}
   for this approximation is again valid only in this limit.
\bibitem{approx}This approximation was used in the past in several papers,
see e.g. Z.-Z.~Chen, R.~L\"u, and B.-F. Zhu, 
Phys.~Rev.~B \textbf{71}, 165324 (2005).
\bibitem{ac_current_Ng} T.~K.~Ng,
   Phys.~Rev.~Lett. \textbf{76}, 487 (1996).
\bibitem{elph_mitra} A.~Mitra, I.~Aleiner, and A.~J.~Millis,
   Phys.~Rev.~B \textbf{69}, 245302 (2004).
\end{thebibliography}
\end{document}